\documentstyle[]{article}

\catcode`\@=11	

\def\@aabuffer{}
\def\author #1{\expandafter\def\expandafter\@aabuffer\expandafter
{\@aabuffer \small\rm      #1\relax \par}}
\def\address#1{\expandafter\def\expandafter\@aabuffer\expandafter
{\@aabuffer \small\it #1\relax \par\vspace{1em}}}

\def\maketitle{
\begin{center}
   {\bf \@title \par}
   \vskip 2em                      
   \@aabuffer\relax
\end{center} \par
\gdef\@aabuffer{}
}

\def\abstracts#1{
\begin{center}
{\begin{minipage}{4.2truein}
                 \footnotesize
                 \parindent=0pt #1\par
                 \end{minipage}}\end{center}
                 \vskip 2em \par}

\fussy
\def\section{\@startsection {section}{1}{\z@}{-3.5ex plus -1ex minus
    -.2ex}{2.3ex plus .2ex}{\bf }}
\def\subsection{\@startsection{subsection}{2}{\z@}{-3.25ex plus -1ex minus
   -.2ex}{1.5ex plus .2ex}{\it }}

\def\@makefnmark{{$\!^{\@thefnmark}$}}

\renewenvironment{thebibliography}[1]
	{\begin{list}{\arabic{enumi}.}
	{\usecounter{enumi}\setlength{\parsep}{0pt}
	\setlength{\itemsep}{0pt}
	\settowidth
	{\labelwidth}{#1.}\sloppy}}{\end{list}}
\def\newblock{} 

\newcounter{arabiclistc}

\def\@citex[#1]#2{\if@filesw\immediate\write\@auxout
	{\string\citation{#2}}\fi
\def\@citea{}\@cite{\@for\@citeb:=#2\do
	{\@citea\def\@citea{,}\@ifundefined
	{b@\@citeb}{{\bf ?}\@warning
	{Citation `\@citeb' on page \thepage \space undefined}}
	{\csname b@\@citeb\endcsname}}}{#1}}

\newif\if@cghi
\def\cite{\@cghitrue\@ifnextchar [{\@tempswatrue
        \@citex}{\@tempswafalse\@citex[]}}
\def\citelow{\@cghifalse\@ifnextchar [{\@tempswatrue
        \@citex}{\@tempswafalse\@citex[]}}
\def\@cite#1#2{{$\null^{#1}$\if@tempswa\typeout
        {IJCGA warning: optional citation argument
        ignored: `#2'} \fi}}

\def\baselinestretch{1.0}
\ifx\selectfont\undefined
\@normalsize\else\let\glb@currsize=\relax\selectfont
\fi

\ifx\selectfont\undefined
\def\@singlespacing{
\def\baselinestretch{1}\ifx\@currsize\normalsize\@normalsize\else\@currsize\fi}
\else
\def\@singlespacing{\def\baselinestretch{1}\let\glb@currsize=\relax\selectfont}\fi

\long\def\@makecaption#1#2{
   \vskip 10pt
   \setbox\@tempboxa\hbox{\footnotesize\rm #1: #2}
   \ifdim \wd\@tempboxa >\hsize   		
       \leftskip 0pt plus 1fil
       \rightskip 0pt plus -1fil
       \parfillskip 0pt plus 2fil
       \footnotesize\rm #1: #2\par   	
     \else                        		
       \hbox to\hsize{\hfil\box\@tempboxa\hfil}
   \fi}

\newcommand{\mathR}{{\rm I\! R}}           

\begin{document}

\title{STRUCTURAL ISSUES IN QUANTUM GRAVITY\footnote{Plenary session
lecture given at the GR14 conference, Florence, August 1995.
Paper completed October 1995}}

\author{C.J. Isham
\footnote{email: c.isham@ic.ac.uk}}

\address{Blackett Laboratory, Imperial College of Science,
Technology and Medicine,\\ South Kensington, London SW7 2BZ}

\maketitle\abstracts{
A discursive, non-technical, analysis is made of some of the basic
issues that arise in almost any approach to quantum gravity, and of
how these issues stand in relation to recent developments in the
field. Specific topics include the applicability of the conceptual
and mathematical structures of both classical general relativity and
standard quantum theory. This discussion is preceded by a short
history of the last twenty-five years of research in quantum
gravity, and concludes with speculations on what a future
theory might look like.  }

\section{Introduction}
\subsection{Some Crucial Questions in Quantum Gravity}
In this lecture I wish to reflect on certain fundamental issues that
can be expected to arise in almost all approaches to quantum
gravity. As such, the talk is rather non-mathematical in nature---in
particular, it is {\em not\/} meant to be a technical review of the
who-has-been-doing-what-since-GR13 type: the subject has developed
in too many different ways in recent years to make this option
either feasible or desirable.

	The presentation is focussed around the following {\em prima
facie\/} questions:
\begin{enumerate}
\item Why are we interested in quantum gravity at all?  In the
past, different researchers have had significantly different
motivations for their work---and this has had a strong influence on
the technical developments of the subject.

\item What are the basic ways of trying to construct a quantum
theory of gravity? For example, can general relativity be regarded
as `just another field theory' to be quantised in a more-or-less
standard way, or does its basic structure demand something quite
different?

\item Are the current technical and conceptual formulations of
general relativity and quantum theory appropriate for the task of
constructing a quantum theory of gravity? Specifically:
	\begin{itemize}

	\item {Is the view of spacetime afforded by general relativity
		adequate at the quantum level?
		In particular, how justified is (i) the concept of a
 		`{\em spacetime point}'; (ii) the assumption that the
		set $\cal M$ of such points has the cardinality of the {\em
		continuum\/}; and (iii) giving this continuum set the
		additional structure of a {\em 	differentiable manifold\/}.

		For example, in most discussions of quantum gravity a central
		role is played by the group ${\rm Diff}{\cal M}$ of spacetime
		diffeomorphisms---which, of course, only makes sense if
		$\cal M$ really is a smooth manifold.
		}

	\item Are the technical formalism and conceptual framework of
		present-day quantum theory adequate for constructing a fully
		coherent theory of quantum gravity? In particular, can they handle
		the idea that `spacetime itself' might have quantum properties in
		addition to those of the metric and other fields that it carries?

\end{itemize}

\end{enumerate}
Note that if our existing views on spacetime and/or quantum theory
are {\em not\/} adequate, the question then arises of the extent to which
they are {\em relevant\/} to research in quantum gravity. Put
slightly differently, are our current ideas about spacetime and
quantum theory of fundamental validity, or are they only heuristic
approximations to something deeper? And, if the latter is true, how {\em
iconoclastic\/} do our research programmes need to be?

	Of course, these are not the only significant issues in quantum
gravity. For example, at a practical level it is important to find
viable {\em perturbative\/} techniques for extracting answers to
physically interesting questions. This is a major challenge to both
superstring theory and the Ashtekar canonical-quantisation
programme: two of the most promising current approaches to quantum
gravity proper.

	Another question---coming directly from the successful
development of superstring theory---concerns the precise role of
supersymmetry in a theory of quantum gravity. Note that
supersymmetry is of considerable significance in modern unified
theories of the {\em non}-gravitational forces, particularly in
regard to having the running coupling constants all meet at a single
unification point.  Thus we are led naturally to one of the central
questions of quantum gravity research: will a consistent theory
necessarily unite all the fundamental forces---as is suggested by
superstring theory---or is it possible to construct a quantum theory
of the gravitational field alone---as is suggested by the Ashtekar
programme?

\subsection{The Peculiar Nature of Research in Quantum Gravity}
Before discussing any of these matters in detail, it is prudent to
point out that the subject of quantum gravity has some distinctly
peculiar features when viewed from the standpoint of most other
branches of theoretical physics. At the end of the day, theoretical
physics is supposed to be about the way things actually `are' in the
physical world: a situation that is reflected in the diagram
\begin{equation}
	\mbox{theory }\longleftrightarrow\makebox{ concepts }
	\longleftrightarrow\makebox{ facts}		\label{theory-concepts-facts}
\end{equation}
in which the theoretical ({\em i.e.}, mathematical) components are
linked to the physical data (the `facts') via a conceptual framework
that depends to some extent on the subject area concerned. Of
course, there is more to this than meets the eye: in particular, it
is well understood these days that (i) the conceptual framework we
use to analyse the world is partly determined by our prior ideas
about the factual content of that world; and (ii) so-called `facts' are not
just bare, `given' data but---in their very identification and
isolation as `facts'---already presuppose a certain conceptual
framework for analysing the world. However, with these caveats in
mind, the simple diagram above does capture certain crucial aspects
of how theoretical physicists view their professional activities.

	The feature of quantum gravity that challenges its very right to
be considered as a genuine branch of theoretical physics is the
singular absence of any observed property of the world that can be
identified {\em unequivocally\/} as the result of some interplay
between general relativity and quantum theory.  This problem stems
from the fact that the natural Planck length---defined using
dimensional analysis as $L_P:=(G\hbar/c^3)^{1\over2}$---has the
extremely small value of approximately $10^{-35}$m; equivalently,
the associated Planck energy $E_P$ has a value $10^{28}$eV, which is
well beyond the range of any foreseeable laboratory-based
experiments. Indeed, this simple dimensional argument suggests
strongly that the only physical regime where effects of quantum
gravity might be studied directly is in the immediate post big-bang
era of the universe---which is not the easiest thing to probe
experimentally.

	This lack of obvious data means that the right hand side of
diagram (\ref{theory-concepts-facts}) is missing, and the shortened
picture
\begin{equation}
\mbox{theory }\longleftrightarrow\mbox{ concepts}	\label{theory-concepts}
\end{equation}
has generated an overall research effort that is distinctly lopsided
when compared to mainstream areas of physics.  In practice, most
research in quantum gravity has been based on various {\em prima
facie\/} views about what the theory {\em should\/} look
like---these being grounded partly on the philosophical prejudices
of the researcher concerned, and partly on the existence of
mathematical techniques that have been successful in what are
deemed, perhaps erroneously, to be closely related areas of
theoretical physics, such as---for example---non-abelian gauge
theories. This procedure has lent a curious flavour to the whole
field of quantum gravity.

	In regard to the lack of experimental data that could act as
a constraint, the situation resembles the one that---until
relatively recently---faced those interested in foundational
problems of quantum theory proper. In this light, it is curious that
there has been such a sparsity of formal interactions between
workers in quantum gravity, and workers in the foundations of
quantum theory itself. This seems all the more remarkable when one
recalls that some of the most basic problems that confront quantum
cosmology are the same as those that have plagued the foundations of
quantum theory in general. I am thinking in particular of the
measurement problem, the meaning of probability, and the general
issue of quantum entanglement in a closed system---in our case, the
universe in its entirety.

\section{Preliminary Remarks}
\subsection{Motivations for Studying Quantum Gravity}
It is clear from the above that, strictly speaking, quantum gravity
cannot be regarded as a standard scientific research programme,
lacking as it does any well-established body of `facts' against
which putative theories can be verified or falsified in the
traditional way. This does not mean there are no good reasons for
studying the subject---there are---but they tend to be of a
different type from those in all other branches of physics. Of
course, some of the motivating factors do refer to potential
observations or experiments---particularly in the area of
cosmology---but most are of a more internal nature: for example, the
search for mathematical consistency, the desire for a unified theory
of all the forces, or the implementation of various
quasi-philosophical views on the nature of space and time. It is
important to appreciate these motivations in order to understand
what people have done in the past, and to be able to judge if they
succeeded in their endeavours: to be adjudged `successful' a theory
must either point beyond itself to new or existing `facts' in the
world, or else achieve some of its own internal goals.

	It is useful pedagogically to classify research programmes in
quantum gravity according to whether they originated in the
community of elementary particle physicists and quantum field
theorists, or in the community of those who work primarily in
general relativity. This divide typically affects both the goals of
research and the techniques employed.

\subsubsection{A.\ Motivations from the perspective of elementary
particle physics and quantum field theory}
\begin{enumerate}
\item Matter is built from elementary particles that {\em are\/}
described in quantum theoretical terms and that certainly interact
with each other gravitationally. Hence it is necessary to say {\em
something\/} about the interface between quantum theory and general
relativity, even if it is only to claim that, `for all practical
purposes', the subject can be ignored (see below).

\item Relativistic quantum field theory might only make proper sense
if gravity is included from the outset. In particular, the
short-distance divergences present in most such theories---including
those that are renormalisable, but not truly finite---might be
removed by a fundamental cut-off at the Planck energy.  Superstring
theory is arguably the latest claimant to implement this idea.

\item A related claim is that general relativity is a necessary
ingredient in any fully-consistent theory of the unification of the
{\em non}-gravitational forces of nature ({\em i.e.}, the
electromagnetic, the weak, and the strong forces). The opposing
positions taken towards the {\em converse\/} claim---that a consistent
quantum theory of gravity will necessarily include the other
fundamental forces---is one of the most striking differences between
superstring theory and the canonical quantum gravity programme.
\end{enumerate}

\subsubsection{B.\ Motivations from the perspective of a general relativist}
\begin{enumerate}
\item Spacetime singularities arise inevitably in
general relativity if the energy-momentum tensor satisfies
certain---physically well-motivated---positivity conditions. It has
long been hoped that the prediction of such pathological behaviour
can be removed by the correct introduction of quantum effects.

\item Ever since Hawking's discovery\cite{Haw75} of the
quantum-induced radiation by a black-hole, a major reason for
studying quantum gravity has been to understand the end state of
gravitationally collapsing matter.

\item {Quantum gravity should play a vital role in the physics of
the very early universe. Possible applications include:
	\begin{enumerate}
	\item understanding the very origin of the universe;

	\item finding an explanation of why spacetime has a macroscopic
	dimension of four (this does not exclude a Kaluza-Klein type higher
	dimension at Planckian scales);

	\item accounting for the origin of the inflationary evolution
	that is felt by many cosmologists to describe
	the universe as it expanded from the initial big bang.
	\end{enumerate}
	}
\end{enumerate}

	In addition to the above---relatively pragmatic---reasons for
studying quantum gravity there remains what is, for many, the most
alluring motivation of all. Namely, a consistent theory of quantum
gravity may require a radical revision of our most fundamental
concepts of space, time and substance. It was John Wheeler who first
most clearly and consistently expounded this thesis over thirty
years ago, and it is one that has fascinated generations of
theoretical physicists ever since.

\subsection{Approaches to Quantum Gravity}
The differing weights placed by individual researchers on the
various motivations for studying quantum gravity have resulted in a
plethora of views on how the subject should be tackled. As a
consequence, we are far from having an `axiomatic' framework,
or---indeed---even a broad consensus on what to strive for beyond
the minimal requirement that the theory should reproduce classical
general relativity and normal quantum theory in the appropriate
domains---usually taken to be all physical regimes well away from
those characterised by the Planck length.

\subsubsection{A.\ Can quantum gravity be avoided?}
Perhaps there is no need for a quantum theory of
gravity at all. In this context, we note the following.
\medskip

	1. The argument is sometimes put forward that the Planck length
$L_P:=(G\hbar/c^3)^{\frac12}\simeq 10^{-35}\mbox{m}$ is so small that
there is no need to worry about quantum gravity except, perhaps, in
recherch\'e considerations of the extremely early universe---{\em
i.e.}, within a Planck time ($\simeq 10^{-42}\mbox{s}$) of the
big-bang. However:
\begin{itemize}
\item Such a claim is only really meaningful if a theory exists
within whose framework genuine perturbative expansions in $L/L_P$
can be performed, where $L$ is the length scale at which the system
is probed: one can then legitimately argue that quantum effects are
ignorable if $L/L_P\ll 1$. So we must try to find a viable theory,
even if we promptly declare it to be irrelevant for anything other
than the physics of the very early universe.

	\item The argument concerning the size of $L_P$ neglects the
possibility of {\em non\/}-perturbative effects---an idea that has
often been associated with the claim that quantum gravity
produces an intrinsic cut-off in quantum field theory.
\end{itemize}
\smallskip

	2. A somewhat different view is that it is manifestly {\em
wrong\/} to attempt to `quantise' the gravitational field in an
active sense
\footnote{By `active' quantisation I mean a diorthotic
scheme in which one starts with a classical system
to which some quantisation algorithm is applied. This can be
contrasted with approaches in which a quantum theory is defined in
an intrinsic way---perhaps as a representation of some group or
algebra---with no prior reference to a classical system that is
being `quantised'.}. The reasons advanced in support of this thesis
include the following.
\begin{itemize}
\item The metric tensor is not a `fundamental' field
in physics, but rather a phenomenological description of
gravitational effects that applies only in realms well away from
those characterised by the Planck scale. One example is superstring
theory, in which the basic quantum entities are far removed from
those in classical general relativity.  Another---somewhat
different---example is Jacobson's recent re-derivation of the Einstein
field equations as an equation of state\cite{Jac95},
which---presumably---it would be no more appropriate to `quantise'
than it would the equations of fluid dynamics
\footnote{In 1971, I took part in a public debate with John Stachel
in which he challenged me on this very issue. As a keen young
quantum field theorist at the time, I replied that I was delighted
to quantise everything in sight. These days I would be more
cautious!}.

	\item The gravitational field is concerned with the structure of
space and time---and these are, {\em par excellence\/}, fundamentally
classical in nature and mode of functioning.
	\end{itemize}

	3. If justified, the last position raises acutely the question
of how matter---which presumably {\em is\/} subject to the laws of
quantum theory---is to be incorporated in the scheme. Discussion of
this issue has largely focussed on the posited equations for the
`semi-classical' spacetime metric $\gamma$,
\begin{equation}
	G_{\mu\nu}(\gamma)=
		\langle\psi|T_{\mu\nu}(g,\widehat{\phi})|\psi\rangle \label{G=<T>}
\end{equation}
where $|\psi\rangle$ is some state in the Hilbert space of the
quantised matter variables $\phi$. In this context, we note the
following:
\begin{itemize}
\item In the case of electromagnetism, the well-known analysis by Bohr
and Rosenfeld\cite{BR33} of the analogue of Eq.\ (\ref{G=<T>})
concluded that the electromagnetic field {\em had\/} to be quantised
to be consistent with the quantised nature of the matter to
which it couples. However---as Rosenfeld himself pointed
out\cite{Ros63}---the analogous argument for general relativity does
not go through and---in spite of much discussion since then (for example,
see Page and Geilker\cite{PG81})---there is arguably still no definitive
proof that general relativity {\em has\/} to be quantised in some way.

	\item The right hand side of Eq.\ (\ref{G=<T>}) generates a
number of technical problems. For example, the expectation value is
ultra-violet divergent, and regularisation methods only yield an
unambiguous expression when the spacetime metric $\gamma$ is static
or stationary---but there is no reason why a semi-classical metric
should have this property. In addition, there have been many
arguments implying that solutions to Eq.\ (\ref{G=<T>}) are likely
to be unstable against small perturbations
and---therefore---physically unacceptable.

	\item It is not clear how the state $|\psi\rangle$ should be
chosen. In addition, if $|\psi_1\rangle$ and $|\psi_2\rangle$ are
associated with a pair of solutions $\gamma_1$ and $\gamma_2$ to
Eq.\ (\ref{G=<T>}), there is no obvious connection between
$\gamma_1$ and $\gamma_2$ and any solution associated with a linear
combination of $|\psi_1\rangle$ and $|\psi_2\rangle$.  Thus the
quantum sector of the theory has curious non-linear features, and these
generate many new problems of both a technical and a conceptual
nature.
\end{itemize}

\subsubsection{B.\ The four types of approach to quantum gravity}
The four major categories in which existing approaches to
quantum gravity can be classified are as follows.

\medskip\noindent
1. {\em Quantise general relativity}. This means trying to construct
an algorithm for actively quantising the metric tensor regarded as a
special type of field. In practice, the techniques that have been
adopted fall into two classes: (i) those based on a spacetime
approach to quantum field theory---in which the operator fields are
defined on a four-dimensional manifold representing spacetime; and
(ii) those based on a canonical approach---in which the operator
fields are defined on a three-dimensional manifold representing
physical space.

\medskip\noindent
2. {\em `General-relativise' quantum theory}. This means trying to
adapt standard quantum theory to the needs of classical general
relativity. Most work in this area has been in the context of
quantising a matter field that propagates on a fixed, background
spacetime $({\cal M},\gamma)$, where $\cal M$ denotes the manifold,
and $\gamma$ is the spacetime metric.

\medskip\noindent
3. {\em General relativity is the low-energy limit of a
quantum theory of something quite different}.\ The most notable
example of this type is the theory of closed superstrings.

\medskip\noindent
4. {\em Start ab initio with a radical new theory}.\ The implication
is that both classical general relativity and standard quantum
theory `emerge' from a deeper theory that involves a radical
reappraisal of the concepts of space, time, and substance.

\subsection{The Problem of Causality and Time}
Until the onset of the superstring programme, most work in quantum
gravity fell into the first category of `quantising' general
relativity.  However, approaches of this type inevitably encounter
the infamous `problem of time' that lies at the heart of many of the
deepest conceptual issues in quantum gravity. For this reason, I
shall begin by briefly reviewing this problem before discussing the
various specific approaches to quantum gravity.

\subsubsection{A.\ The problem of time from a spacetime perspective}
In the context of spacetime-oriented approaches to quantum theory,
the problem of time and causality is easy to state: the causal
structure of spacetime depends on the metric tensor
$\gamma$---hence, if this is subject to quantum fluctuations, so is
the causal structure. Similarly, if the metric is only a
coarse-grained, phenomenological construct of some type, then so is
the causal structure.

	This situation poses severe technical problems since standard
quantum field theory presupposes a fixed causal structure. For
example, a quantum scalar field $\widehat\phi(X)$ is normally
required to satisfy the microcausal commutation relations
\begin{equation}
	[\,\widehat\phi(X),\,\widehat\phi(Y)\,]=0	\label{microcausality}
\end{equation}
whenever the spacetime points $X$ and $Y$ are spacelike separated.
However, the latter condition has no meaning if the spacetime metric
is quantised or phenomenological. In the former case, the most
likely scenario is that the right hand side of Eq.\
(\ref{microcausality}) never vanishes, thereby removing one of the
foundations of conventional quantum field theory\cite{FH87}.
Replacing operator fields with $C^*$-algebras does not help---in so
far as they have microcausal commutation properties---and neither
does the use of functional integrals if the problem of their
definition is taken at all seriously. In practice, the techniques
that have been used to address the problem of time fall into one of
the following categories:
\begin{enumerate}

\item {Use a fixed background metric $\eta$---often
chosen to be that of Minkowski spacetime---to define a fiducial
causal structure with respect to which standard quantum field
theoretical techniques can be employed. When applied to the
gravitational field itself, $\gamma_{\mu\nu}(X)$, this usually
involves writing
\begin{equation}
	\gamma_{\mu\nu}(X)=\eta_{\mu\nu}+\kappa h_{\mu\nu}(X)
									\label{g=eta+h}
\end{equation}
and then regarding $h_{\mu\nu}(X)$ as the physical, `graviton' field
(here, $\kappa^2=8\pi G/c^2$ where $G$ is Newton's constant). The
use of Eq.\ (\ref{g=eta+h}) strongly suggests a perturbative
approach in which quantum gravity is seen as a theory of small
quantum fluctuations around a background spacetime. The problems
that arise include:
\begin{itemize}
	\item finding a well-defined mathematical scheme
	(the obvious techniques give a non-renormalisable theory---see below);

	\item knowing how to handle backgrounds that are other than
Minkowski spacetime, both metrically and topologically---for
example, to discuss cosmological issues;

	\item understanding if---and how---the different possible backgrounds fit
	together into a single quantum scheme, and what becomes of the notion of
	`causality' in such a scheme.
	\end{itemize}
	}

\item Start with a formalism in which the spacetime metric has a
Riemannian (rather than Lorentzian) signature, and then worry at the
end about making an `analytical continuation' back to physical
spacetime.  Most work of this type has involved---rather
heuristic---functional-integral approaches to quantisation.

\item Forget about spacetime methods altogether and adopt a
canonical approach to general relativity in which the basic
ingredients are geometrical fields on a three-dimensional manifold.
The problem then is to reconstruct some type of---possibly, only
approximate---spatio-temporal picture within which the quantum
calculations can be interpreted.
\end{enumerate}

	The absence of any fundamental causal structure also
raises important conceptual issues. For example, the standard
interpretation of quantum theory places much weight on the role of
measurements made by an `observer'. But the simplest model for an
observer is a time-like curve, and the notion of `time-like' is
dependent on the spacetime metric. Thus what it is to be an
observer also becomes quantised---or phenomenological, as the case
may be---which renders the standard interpretation distinctly
problematic
\footnote{I am grateful to Steve Weinstein for emphasising this
point to me, and for discussions on its significance.}.

\subsubsection{B.\ The problem of time in canonical quantisation}
The canonical approach to quantum gravity starts with a reference
foliation of spacetime that is used to define the appropriate
canonical variables. These are the $3$-metric $g_{ab}(x)$ on a
spatial manifold $\Sigma$, and a canonical conjugate $p^{ab}(x)$
that---from a spacetime perspective---is related to the extrinsic
curvature of $\Sigma$ as embedded in the four-dimensional spacetime.
However, these variables are not independent, but satisfy the
constraints
\begin{eqnarray}
	{\cal H}_{a}(x)&=& 0				\label{Ha=0}		\\
	{\cal H}_\perp(x)&=& 0				\label{Hperp=0}
\end{eqnarray}
where
\footnote{The `$|$' sign denotes covariant differentiation
using the Christoffel symbol of the $3$-metric.}
\begin{eqnarray}
    {\cal H}_a(x) &:= &-2{p_a{}^b}_{|b}(x)             \label{Def:Ha}\\
    {\cal H}_\perp(x) &:=& \kappa^2{\cal G}_{ab\,cd}(x)\,
		p^{ab}(x)\,p^{cd}(x) -
			{|g|^{\frac12}(x)\over\kappa^2}\,R(x)     \label{Def:Hperp}
\end{eqnarray}
in which $R(x)$ denotes the curvature scalar of the $3$-metric
$g_{ab}(x)$, and where the `DeWitt supermetric' on the space of
three-metrics\cite{DeW67a} is given by
\begin{equation}
    {\cal G}_{ab\,cd}(x):=
        \frac12|g|^{-\frac12}(x)\big(g_{ac}(x)g_{bd}(x)+g_{bc}(x)g_{ad}(x)
           -g_{ab}(x)g_{cd}(x)\big).             \label{Def:DeW-metric}
\end{equation}
The functions ${\cal H}_a$ and ${\cal H}_\perp$ of the canonical
variables $(g,p)$ play a key role, centered on the fact that their
Poisson bracket algebra
\begin{eqnarray}
\big\{{\cal H}_a(x),{\cal H}_b(x')\big\} &=&
	-{\cal H}_b(x)\,\partial_a^{x'}\delta(x,x') +
     	{\cal H}_a(x')\,\partial_b^x\delta(x,x')       	\label{PB:HaHb}\\
\big\{{\cal H}_a(x),{\cal H}_\perp(x')\big\} &=&
        {\cal H}_\perp(x)\,\partial_a^x\delta(x,x')        \label{PB:HaH}\\
\big\{{\cal H}_\perp(x),{\cal H}_\perp(x')\big\} &=&
	g^{ab}(x)\,{\cal H}_a(x)\,\partial^{x'}_b\delta(x,x')-  \nonumber\\
 &\ &\ \ \ \ g^{ab}(x')\,{\cal H}_a(x')\,\partial_b^x\delta(x,x')
															\label{PB:HH}
\end{eqnarray}
is that of the spacetime diffeomorphism group projected along, and
normal to, the spacelike hypersurfaces.

	The way in which the problem of time appears depends very much
on the approach taken to quantising this classical canonical system.
One possibility is to (i) impose a gauge for the invariance
associated with the algebra Eqs.\ (\ref{PB:HaHb}--\ref{PB:HH}); (ii)
solve the constraints Eqs.\ (\ref{Ha=0}--\ref{Hperp=0}) classically;
and (iii) quantise the resulting `true' canonical system in a
standard way.  The final equations are intractable in anything other
than a perturbative sense, where they promptly succumb to
ultraviolet divergences. However, if the formalism could be given a
proper mathematical meaning, the problem of time would involve
relating the different choices of gauge and associated classical
notions of time. Simple model calculations suggest that this is far
from trivial.

	Most approaches to canonical quantum gravity do not
proceed in this way. Instead, the full set of fields $(g_{ab}(x),
p^{cd}(x))$ is quantised via the canonical commutation relations
\begin{eqnarray}
    [\,\widehat g_{ab}(x),\widehat g_{cd}(x')\,]  &=& 0    \label{CR:gg}\\
    {[}\,\widehat p^{ab}(x),\widehat p^{cd}(x')\,] &=& 0    \label{CR:pp}\\
    {[}\,\widehat g_{ab}(x),\widehat p^{cd}(x')\,] &=&
          i\hbar\,\delta^c_{(a}\delta^d_{b)}\,\delta(x,x') 	\label{CR:gp}
\end{eqnarray}
of operators defined on the $3$-manifold $\Sigma$. Following Dirac,
the constraints are interpreted as constraints on the allowed state
vectors $\Psi$, so that $\widehat {\cal H}_a(x)\Psi=0=\widehat{\cal
H}_\perp(x)$ for all $x\in\Sigma$. In particular, on choosing the
states as functions of the three-geometry $g$---and with operator
representatives $(\widehat g_{ab}(x)\Psi)[g]:=g_{ab}(x)\Psi[g]$ and
$(\widehat p^{cd}(x)\Psi)[g]:=-i\hbar\delta\Psi[g]/\delta
g_{ab}(x)$---the constraints $\widehat{\cal H}_a\Psi=0$
imply that $\Psi[g]$ is constant under changes of $g$ induced by
infinitesimal diffeomorphisms of the spatial 3-manifold $\Sigma$.

	The crucial constraint is $\widehat{\cal H}_\perp(x)\Psi=0$,
which---in this particular representation of states and
canonical operators---becomes the famous Wheeler-DeWitt equation
\begin{equation}
  -\hbar^2\kappa^2{\cal G}_{ab\,cd}(x)
	{\delta^2\Psi[g]\over\delta g_{ab}(x)\,\delta g_{cd}(x)}
		-{|g|^{\frac12}(x)\over\kappa^2}\,R(x)\Psi[g]=0     \label{WDE}
\end{equation}
where ${\cal G}_{ab\,cd}$ is the DeWitt metric defined in
Eq.\ (\ref{Def:DeW-metric}).

	The most obvious manifestation of the problem of time is that
the Wheeler-DeWitt equation Eq.\ (\ref{WDE}) makes no reference to time,
and yet this is normally regarded as the crucial `dynamical'
equation
\footnote{Essentially because in the classical theory---when
viewed from a spacetime perspective---${\cal H}_\perp$ is
associated with the canonical generators of displacements in
time-like directions.} of the theory!  This situation is usually
understood to mean that `time' has to be reintroduced as the values
of special {\em physical\/} entities in the theory---either
gravitational or material---with which the values of other physical
quantities are to be correlated. It is a major unsolved problem
whether (i) this can be done at all in an exact way; and (ii) if it
can, how the results of two different such choices compare with each
other, and how this can be related to concepts of a more
spatio-temporal nature.

	Note that even the starting canonical commutation relations
Eqs.\ (\ref{CR:gg}--\ref{CR:gp}) are suspect. For example, the
vanishing of a commutator like Eq.\ (\ref{CR:gg}) would normally
reflect the fact that the points $x$ and $x'$ are `spatially
separated'.  But what does this mean in a theory with no background
casual structure? Questions of this type have led many people to
question the whole canonical approach to quantum gravity, and have
generated searches for a new---essentially `timeless'---approach to
quantum theory itself. However, the problem of time is very complex
and is still the subject of much debate. Two recent extensive
reviews are by Kucha{\v r}\cite{Kuc92a} and Isham\cite{Ish93}.

\section{A Brief History of Quantum Gravity}
Rather than just summarising recent developments in quantum gravity I
would like to start by presenting a short history
\footnote{Please note that had I tried to give full references,
the bibliography would have consumed the entire page allowance for
this article!  Therefore, in several places I have had to be
satisfied with merely citing review papers containing full
bibliographies for specific subjects.} of the subject as it has
developed over the last twenty-five years: if we are interested in
speculating on where quantum gravity is going, it is not
unreasonable to reflect first on where it has come from!

\subsection{The Situation Twenty-Five Years Ago}
Let me begin by recalling the status of quantum gravity studies in
the year 1970 ({\em i.e.,} twenty-five years before this present
GR14 conference) in particular, the state
of quantum gravity proper, and the way elementary particle physics
and general quantum field theory were viewed at that time.

\subsubsection{A. Quantum gravity before 1970}

1.\ The canonical analysis of classical general relativity was
well understood by this time. The pioneering work of
Dirac\cite{Dir58b} had been developed by many people, with one line
of research culminating in the definitive treatment by Arnowitt,
Deser and Misner\cite{ADM62} of how to isolate the physical degrees
of freedom in classical general relativity
\footnote{A fairly comprehensive bibliography of papers on canonical
general relativity can be found in my review paper\cite{Ish92} on
the problem of time, and in a review by Kucha{\v r}\cite{Kuc93}.}.
The classical constraint algebra Eqs.\ (\ref{PB:HaHb}--\ref{PB:HH}) was
also well-known, and its broad implications for the quantum theory
were understood. In particular, the Wheeler-DeWitt equation
Eq.\ (\ref{WDE}) had been written down (by Wheeler\cite{Whe64,Whe68} and
DeWitt\cite{DeW67a}, of course).

\medskip\noindent
2.\ The first studies of quantum cosmology had been made.  In
particular, DeWitt\cite{DeW67a} and Misner\cite{Mis69a} had
introduced the idea of {\em minisuperspace\/} quantisation: a
truncation of the gravitational field to just a few
degrees of freedom so that the---rather intractable---functional
differential equation Eq.\ (\ref{WDE}) becomes a partial differential
equation in a finite number of variables, which one can at least
contemplate attempting to solve exactly.

\medskip\noindent
3.\ There was a fair appreciation of many of the conceptual
problems of quantum gravity. In particular:
\begin{itemize}
	\item the instrumentalist concepts central to the Copenhagen
interpretation of quantum theory were understood to be inappropriate
in the context of quantum cosmology (recall that the many-worlds
interpretation dates back to papers by Everett\cite{Eve57} and
Wheeler\cite{Whe57} that were published in 1957!);

	\item there had been a preliminary analysis of the problem of
time, especially in the context of canonical quantum gravity;

	\item doubts had been expressed about the operational meaning of a
`space-time point' in quantum gravity---for example, the
twistor programme has its genesis in this era\cite{Pen67}.
\end{itemize}

\medskip\noindent
4.\ A number of studies had been made of the spacetime approach to
quantum gravity centered around the expansion Eq.\ (\ref{g=eta+h}). When
substituted into the Einstein-Hilbert action $S=\int d^4X
|\gamma|^{\frac12} R(\gamma)$, this gives a bilinear term that
describes massless spin-2 gravitons, plus a series of higher-order
vertices describing graviton-graviton interactions. The Feynman
rules for this system were well understood, including
the need to introduce `ghost' particles to allow for the
effects of non-physical graviton modes propagating in internal
loops\cite{Fey63,DeW65,DeW67a,DeW67b,Man68b,FP67}. And---most
importantly---the theory was widely expected to be
non-renormalisable, although only a simple power-counting argument
was available at that time.

\subsubsection{B.\ Elementary particle physics and quantum
field theory before 1970}
The attitude in the 1960s towards quantum field theory was very
different from that of today. With the exception of quantum
electrodynamics, quantum field theory was poorly rated as a
fundamental way of describing the interactions of elementary
particles.  Instead, this was the era of the S-matrix, the Chew
axioms, Regge poles, and---towards the end of the period---the dual
resonance model and the Veneziano amplitude that led eventually to
string theory.

	In so far as it was invoked at all in strong interaction
physics, quantum field theory was mainly used as a phenomenological
tool to explore the predictions of current algebra, which was
thought to be more fundamental. When quantum field theory was
studied seriously, it was largely in the context of an `axiomatic'
programme---such as the Wightman\cite{SW64} axioms for the $n$-point
functions.

	This general down-playing of quantum field theory influenced the
way quantum gravity developed. In particular---with a few notable
exceptions---physicists trained in particle physics and quantum
field theory were not interested in quantum gravity, and the subject
was left to those whose primary training had been in general
relativity. This imparted a special flavour to much of the work in
that era. In particular, the geometrical aspects of the theory were
often emphasised at the expense of quantum field theoretic
issues---thereby giving rise to a tension that has affected the
subject to this very day.

\subsection{The Highlights of Twenty-Five Years of Quantum Gravity}
My personal choice of the key developments in the last twenty-five
years is as follows.

\medskip\noindent
1. {\em The general renaissance of Lagrangian quantum field
theory}.\ I make no apologies for beginning with the discovery by
t'Hooft in the early 1970s of the renormalisability of quantised
Yang-Mills theory.  Although not directly connected with gravity,
these results had a strong effect on attitudes towards quantum field
theory in general and reawakened a wide interest in the subject. One
spin-off was that many young workers in particle physics became
intrigued by the challenge of applying the new methods to quantum
gravity---a trend that has continued to the present time.

\medskip\smallskip\noindent
2. {\em Black hole radiation}.\ I was present at the Oxford
conference\cite{IPS81} in 1974 at which Hawking announced
his results on black hole radiation\cite{Haw75}, and
I remember well the amazement engendered by his lecture. His seminal
work triggered a series of research programmes that have been of
major interest ever since. For example:

\medskip
(a) The most obvious conclusion at the time was that there is
some remarkable connection between thermodynamics---especially the
concept of entropy---quantum theory, and general relativity. The
challenge of fully elucidating this connection has led to some of
the most intriguing ideas in quantum gravity (see later).

\medskip
(b) Hawking's work generated an intense---and ongoing---interest
in the general problem of defining quantum field theory on a curved
space-time background
\footnote{For a recent overview and bibliography see the paper
by Bob Wald in this volume.}.

\medskip
(c) Hawking's results were quickly rederived using thermal Green's
functions\cite{GP78} which---in normal quantum field theory---are
closely connected with replacing time by an imaginary number whose
value is proportional to the temperature.  This led Hawking to
propose his `Euclidean' quantum gravity programme
\footnote{A convenient recent source for many of the original articles is
Gibbons and Hawking\cite{GH93}.} in which the central role is played
by Riemannian, rather than Lorentzian, metrics (this being the
appropriate curved-space analogue of replacing time $t$ with
$\sqrt{-1}t$). In particular, Hawking proposed to study functional
integrals of the form
\begin{equation}
	Z({\cal M}):=\int{\cal D}\gamma\,
		e^{-\int_{\cal M}|\gamma|^{\frac12}R(\gamma)}		\label{Def:Z(M)}
\end{equation}
where the integral is over all Riemannian metrics
$\gamma$ on a four-manifold $\cal M$. It is not easy
to give a rigorous mathematical meaning to this object but,
nevertheless, the idea has been extremely fertile. For example:
\begin{itemize}
\item Solving the problem of time involves the `analytical
continuation' of manifolds with a Riemannian signature to those whose
signature is Lorentzian. This procedure is of considerable mathematical
interest in its own right.

\item Saddle-point approximations to Eq.\ (\ref{Def:Z(M)})
have been widely used as a gravitational analogue of the
instanton techniques developed in Yang-Mills theory. This has
generated considerable mathematical interest in classical
solutions to the Riemannian version of general relativity.

\item The expression Eq.\ (\ref{Def:Z(M)}) generalises naturally to
include a type of `quantum topology' in which each four-manifold $\cal
M$ contributes with a weight $\chi({\cal M})$ in an expression of the
type
\begin{equation}
Z:=\sum_{\cal M}\chi({\cal M})Z({\cal M}).	\label{Z=sumZ(M)}
\end{equation}

\item If applied to a manifold with a single three-boundary
$\Sigma$, the expression Eq.\ (\ref{Def:Z(M)}) gives rise to a functional
$\Psi[g]$ if the functional integral is taken over all 4-metrics
$\gamma$ on $\cal M$ that are equal to the given 3-metric $g$ on
$\Sigma$.  Furthermore, the functional of $g$ thus defined satisfies
(at least, in a heuristic way) the Wheeler-DeWitt equation
Eq.\ (\ref{WDE}).  This is the basis of the famous
Hartle-Hawking\cite{HH83} `wave-function of the universe' in quantum
cosmology.
\end{itemize}

\bigskip\noindent
3.\ {\em The non-renormalisability of quantum gravity}.\ Around
1973, a number of calculations were performed
\footnote{Full references can be found in reviews written around that time; for
example, in the proceedings of the first two Oxford conferences on
quantum gravity\cite{IPS75,IPS81}.} confirming that perturbative
quantum gravity is indeed non-renormalisable
\footnote{More precisely, it was shown that---in a variety of
matter-plus-gravity systems---the one-loop counter-term is
ultraviolet divergent (in the background-field method, the one-loop
counter-term for pure gravity vanishes for kinematical
reasons).  In 1986, Goroff and Sagnotti\cite{GS86} showed that the
two-loop contribution in pure gravity is also infinite.
Thus---barring a very improbable, fortuitous cancellation of all
higher infinities---the theory is non-renormalisable.}. There have
been three main reactions to this situation, each of which still has
many advocates today:
\begin{itemize}
	\item Continue to use standard perturbative quantum field theory
but change the classical theory of general relativity so
that the quantum theory becomes renormalisable. Examples of
such attempts include (i) adding higher powers of the Riemann
curvature $R^\alpha_{\beta\mu\nu}(\gamma)$ to the action; and (ii)
supergravity (see later).

	\item Keep classical general relativity as it is, but develop
quantisation methods that are intrinsically non-perturbative. The
two best examples of this philosophy are (i) the use of Regge
calculus and techniques based on lattice gauge theory; and
(ii) the Ashtekar programme for canonical quantisation (see below).

	\item Adopt the view that the non-renormalisability of
perturbative quantum gravity is a catastrophic failure that can be
remedied only by developing radically new ideas in the foundations of
physics.  Research programmes of this type include (i) twistor
theory; (ii) non-commutative geometry, as developed by
Connes\cite{Con94}, Dubois-Violette
\footnote{A comprehensive recent review is by Djemai\cite{Dje95}.}
and others; and (iii) a variety of ideas involving discrete models
of space and time
\footnote{A bibliographic review of some of the lesser known schemes
has been written recently by Gibbs\cite{Gib95}.}.
\end{itemize}

\medskip\smallskip\noindent
4.\ {\em Supergravity and superstrings}.\ The development of
supergravity
\footnote{A full bibliography is contained in a definitive review
written in 1980 by van Nieuwenhuizen\cite{vN81}.} (in the mid 1970s)
and superstring theory (mainly from the mid 1980s onwards) has had a
major impact on the way the problem of quantum gravity is viewed.
For example:
\begin{itemize}
\item These theories provide a concrete realisation of the old
hope that quantum gravity necessarily involves a unification with
other fundamental forces in nature.

\item Superstring theory shows clearly how general relativity can
occur as a small part of something else---thereby removing much of
the fundamental significance formerly ascribed to the notions of
space and time. Not unsurprisingly, this radical implication of the
superstring programme tends not to be overwhelmingly popular with
the general relativity community!

\item There is clear evidence in string theory of the existence of
non-local structure at around the Planck length. Indeed, there have
been frequent suggestions that the very notion of `length' becomes
meaningless below this scale---an idea that turns up frequently in
many other approaches to quantum gravity\cite{Pru95,Gar95}.
\end{itemize}

\medskip\smallskip\noindent
5.\ {\em The new canonical variables}.\ A major advance in the
development of the canonical quantisation of gravity occurred in 1986
when Abhay Ashtekar\cite{Ash86} found a set of canonical variables
in terms of which the structure of the constraints is dramatically
simplified. Ever since, there has been a very active programme
to exploit these new variables in both classical
and quantum gravity. Some of the more striking implications include
the following.
\begin{itemize}
\item For the first time, there is real evidence in support of the old idea
that non-perturbative methods must play a key role in constructing a
quantum theory of gravity.

\item The new variables involve complex combinations of the
standard canonical variables. Thus the use of {\em complexified\/}
general relativity moves to the front of the quantum stage.

\item One of the new variables is a spin-connection, which
suggests the use of a gravitational analogue of the
gauge-invariant loop variables introduced by Wilson in Yang-Mills
theory. Seminal work in this area by Rovelli and Smolin\cite{RS90}
has produced many fascinating ideas, including a claim that the
area and volume of space are quantised
\footnote{For full references on this---and other---aspects of
the Ashtekar programme, see the paper by Abhay in this volume.}
in discrete units.
\end{itemize}
\medskip

	This concludes my list of the main highlights of the last
twenty-five years of research. However, this by no means
exhausts the topics on which people have worked. In particular:
\begin{enumerate}
\item There has been a very successful research
programme
\footnote{For a recent review see the lecture notes\cite{Car95a} by
Carlip.} aimed at understanding quantum gravity in $2+1$-dimensional
space-time. This has been particularly valuable for illustrating the
relations between different approaches to quantum gravity. It
also provides a viable platform for analysing some of the many
conceptual problems that plague quantum gravity, in a way that is
free of the---often intractable---mathematical problems that infest
the theory in $3+1$ dimensions.

\item There has been a steady growth of interest in `topological
quantum field theories' (see later).

\item Many people have continued to think long and hard about the
conceptual issues
\footnote{Many of these are discussed at
length in the notes of my 1991 Schladming lectures\cite{Ish92}; see
also the proceedings\cite{AS91} of the 1988 Osgood Hill conference
on conceptual problems in quantum gravity.} in quantum gravity. In
particular, there have been intensive studies of (i) the problem of
time in canonical quantum gravity, and the associated problem of the
nature of physical observables; (ii) the possibility of finding new
interpretations of quantum theory that avoid the instrumentalism of
the standard approaches; and (iii) the possibility that quantum
gravity can solve some of the conceptual problems in normal quantum
theory---in particular, the idea that the `reduction of the state
vector' can be associated with the non-linear nature of general
relativity (see later).
\end{enumerate}

	At this point---having completed my brief historical survey---it
might seem natural to list the research areas that are currently
active.  However, in fact this is hardly necessary since almost every topic
mentioned above is still being pursued in one way or another.
Indeed, reflecting on these topics suggests that---like old
soldiers---ideas in quantum gravity do not die but merely fade
away---in some cases, over quite a long time scale! As far as
quantum gravity proper is concerned, there is currently much
activity in all three of the major approaches: (i) superstring
theory; (ii) the Ashtekar programme; and (iii) the Euclidean quantum
gravity programme. These three programmes complement each other
nicely and enable the special ideas of any one of them to be viewed
in a different perspective by invoking the other two---a feature
that is rather useful in a subject that so singularly lacks any
unequivocal experimental data.

\section{Structural Issues Concerning Space and Time}
I shall focus on four issues concerning the use in quantum gravity
of the picture of spacetime suggested by classical general
relativity. Namely (i) the representation of spacetime as a
$C^\infty$-manifold; (ii) the role of spacetime diffeomorphisms;
(iii) the role of black holes; and (iv) the implications of recent
developments in superstring theory.

\subsection{Is Spacetime a $C^\infty$-Manifold?}
In classical general relativity, the basic mathematical structure is
a pair $({\cal M},\gamma)$ where the smooth differentiable manifold
$\cal M$ represents spacetime, and $\gamma$ is a
Lorentzian-signature metric defined on $\cal M$. Similarly, space is
modeled by a three-dimensional manifold $\Sigma$ equipped with a
Riemannian metric $g$.

	In the context of a quantum theory of gravity, the first crucial
question is whether it is still correct to base everything on an
underlying set $\cal M$ of `spacetime points'. If so, is the correct
mathematical structure still differential geometry, or should a
different---perhaps broader---category like general point-set
topology be used?  Implicit in such questions is the idea that---in
addition to the fields it carries---the structure of `spacetime
itself' may be subject to quantum effects. The broad options seem to
be the following.
\begin{enumerate}
	\item {The theory of quantum gravity requires a fixed set of
spacetime (or, if appropriate, spatial) points equipped with a fixed
topological and differential structure.

	Thus spacetime (or space) itself is the same as in classical
general relativity. This is the view adopted by the canonical
quantum gravity programme. It is also inherent in spacetime oriented
quantisation schemes based on the expansion Eq.\ (\ref{g=eta+h}),
and in simple versions of perturbative superstring theory (see
later).  }

	\item {The theory of quantum gravity requires a fixed set of
spacetime (or spatial) points.  However, the topology and/or
differential structure on this set is subject to quantum effects.

	For example, in the context of canonical quantisation, Wheeler
\cite{Whe68} suggested that large quantum fluctuations in the metric
tensor could induce fluctuations in the spatial topology---what I
shall call `metric-driven' topology changes. Another---but not
necessarily unrelated---possibility is that the topological
structure of space or spacetime is `{\em actively\/} quantised' in
some way; an example in the Euclidean quantum gravity programme is
the sum Eq.\ (\ref{Z=sumZ(M)}) over manifolds $\cal M$. In either
case, the question arises whether the resulting quantum effects
could move from the category of differentiable manifolds to something
more general. If so, one might be cautious about starting with a
formalism ({\em i.e.}, classical general relativity) in which
differential geometry plays such a fundamental role from the outset.  }

	\item {The notion of a spacetime point is not meaningful at a
fundamental level.

	In particular, the language of differential geometry employed in
classical general relativity is a {\em phenomenological\/} tool that
applies only at scales well away from the Planck length or energy.
		}

\end{enumerate}

\subsubsection{A.\ The possibility of asymptotic freedom}
The idea of metric-driven topology change is based on the intuition
that the effects of quantum gravity become more pronounced at
decreasing distances, resulting eventually in a `foamlike' structure
at around the Planck length. However, an alternative view is that
quantum gravity could be {\em asymptotically-free\/}---in which case
the effects become smaller, not larger, as the scale reduces. Under
these circumstances, there would be no metric-driven topology
changes.

	Asymptotic freedom would also mean that semi-classical methods
could give physically useful predictions at very small scales: a
possibility that has been exploited recently by Brandenburger\cite{Bra95}
in a cosmological context.  The idea that gravity might be
asymptotically-free was studied some years ago by Fradkin and
Tseytlin\cite{FT81} in the context of $R+R^2$ theories of gravity.
More recently, Tseytlin\cite{Tse95} has emphasised the importance of
the analogous effect in superstring theory.

\subsubsection{B.\ Spacetime as a phenomenological construct}
The notion of active quantisation of topology raises a number
of fundamental questions that will be addressed in Section
\ref{SSec:QuSpaceTime} in the course of a general discussion of the
applicability of present-day quantum theory. However, my personal
leanings are towards the more iconoclastic view that the concept of
`spacetime' is not a fundamental one at all, but only something
that applies in a `phenomenological' sense when the universe is
not probed too closely. Of course, in modern quantum field theory we
have become accustomed to the idea of phenomenological schemes that
only work with some degree of coarse-graining of the physical world.
However, all existing theories of this type employ a strictly
classical view of the fundamental nature of the manifold of
spacetime (or spatial) points, whereas what is being suggested now
is that spacetime itself is also a concept of strictly limited
applicability.

	The most obvious thing to regard as phenomenological in this
sense is probably the topology or differential structure on a fixed
set of spacetime points. However, the phenomenological status might
extend to the notion of a spacetime point itself.  Note that,
if correct, ideas of this type imply that the first two options
above are definitely incorrect: it is wrong to work with a fixed spacetime
manifold---because that is not a  meaningful concept at a
fundamental level---but it is also wrong to talk about `actively
quantising' such things if there is no such thing to quantise.

\begin{figure}[t]
\begin{picture}(400,330)(0,100)
	\put(0,400){\makebox(180,25){The `ultimate' theory\ \ \qquad
		$\longleftrightarrow$}
			\framebox(100,25){Exact results}}
	\put(230,370){$\Bigg\downarrow$\ \ `coarse-grain'}
	\put(0,320){\makebox(180,25){Phenomenological theory\qquad
		$\longleftrightarrow$}
			\framebox(100,25){Limited results}}
	\put(230,290){$\Bigg\downarrow$\ \ `coarse-grain'}
	\put(0,240){\makebox(180,25){Phenomenological theory\qquad
		$\longleftrightarrow$}
			\framebox(100,25){Limited results}}
	\multiput(233,185)(0,10){5}{\line(0,1){5}}
	\put(230,163){$\Bigg\downarrow$}
	\put(245,200){`coarse-grain'}
\put(0,115){\makebox(175,25){Spacetime $\cal M$ \&
	${\rm Diff}({\cal M})$\qquad $\longleftrightarrow$}
			\framebox(115,25){Algebra of `local regions'}}

\put(50,90){Figure 1.\ The hierarchy of phenomenological theories}
\end{picture}

\end{figure}

	The general idea behind this view of spacetime is sketched in
Figure 1. At the top of the tower is the `ultimate' theory of
physics, whose fundamental categories---we are supposing---do not
include continuum ideas of spacetime or, indeed, ideas of spacetime
points at all. A `phenomenological theory' means a mathematical
structure that replicates only certain coarse-grained features of
the fundamental theory.  This structure may itself be coarse-grained
further, and so on, leading eventually to a mathematical model in
which our conventional ideas of space and time can be recognised.
This could happen in various ways. For example, one result of
coarse-graining might be the idea of a `local region'---not regarded
as a subset of something called `spacetime', but rather as an
emergent concept
\footnote{Of course, this could also exist already in the top level of
the tower as a central ingredient in the fundamental theory.} (like
pressure, for example) in its own right---plus an algebra of such
regions that specifies their theoretical use, and that can be
identified mathematically as the algebra of a certain open
covering\cite{Sor83,Sor91a,Ish90d} of a genuine continuum manifold
$\cal M$. Hence---as long as one keeps to the phenomena appropriate
to this level---it is {\em as if\/} physics is based on the
spacetime manifold $\cal M$, even though this plays no fundamental
role in the `ultimate' theory with which we started.

	Of course, many philosophical---as well as mathematical and
physical---issues are involved in a picture of this type. For
example, the idea of an `ultimate' theory may be meaningless; in
which case the tower in Figure 1 has no upper member at all. And
many different towers may branch off from the same level, thus
raising general issues of realism and instrumentalism. However, the
central idea---that concepts of spacetime point, topology, and
differential structure---have no fundamental status, is one that
could well form an important ingredient in future theories of
quantum gravity.

\subsection{The Group of Spacetime Diffeomorphisms}
When thinking about the role of spacetime $\cal M$ (or space
$\Sigma$) in quantum gravity, one key issue is the status of the
associated group of diffeomorphisms ${\rm Diff}({\cal M})$ (resp.
${\rm Diff}({\Sigma})$) that is such a central feature of the
classical theory of general relativity. There are at least three
ways in which ${\rm Diff}({\cal M})$ or ${\rm Diff}(\Sigma)$ could
appear in the quantum theory: (i) as an {\em exact\/} covariance
group; (ii) as a {\em partial} covariance group; (iii) as a limited
concept associated with a phenomenological view of spacetime (or
space).  Note that the third option flows naturally from the
view of spacetime promulgated above: if spacetime is a
phenomenological concept of limited applicability, then so will be
the diffeomorphisms of the manifold that models spacetime
in this limited sense.  All three potential roles of the
diffeomorphism group merit attention, and each will be discussed
briefly in what follows.

\subsubsection{A.\ Diffeomorphisms as an exact covariance group}
The idea that the group of spacetime diffeomorphisms is an exact
covariance group plays a key role in many existing approaches to
quantum gravity. For example, it is one of the defining properties
of a topological quantum field theory. It also plays a major role in
canonical quantum gravity via the classical Poisson bracket algebra
Eqs.\ (\ref{PB:HaHb})--(\ref{PB:HH}) of the constraint functions
${\cal H}_a$ and ${\cal H}_\perp$ that can be interpreted as the
algebra of spacetime diffeomorphisms projected along, and normal to,
spacelike hypersurfaces.  In approaches to the quantum theory in
which all twelve functions $g_{ab}(x)$ and $p^{cd}(x)$ are quantised
(with the canonical algebra Eqs.\ (\ref{CR:gg})--(\ref{CR:gp})), it
is natural to suppose that Eqs.\ (\ref{PB:HaHb})--(\ref{PB:HH}) are
to be replaced with the analogous commutator algebra of the quantum
operators $\widehat{\cal H}_a$ and $\widehat{\cal H}_\perp$; indeed,
if the classical algebra is {\em not\/} preserved by the quantum
theory there is a great danger of anomalous quantum excitations of
non-physical modes of the gravitational field.

	Of course, even to talk of such things requires the operators
$\widehat{\cal H}_a$ and $\widehat{\cal H}_\perp$ to be defined
rigorously---a task that is highly non-trivial, not least because
this is the point at which ultraviolet divergences are likely to
appear. Indeed, the possibility of addressing this issue properly
only arose fairly recently following significant advances in the
Ashtekar programme, and the issue of anomalies there will clearly be
a crucial one in the next few years; particularly in the
\mbox{supergravity} version where extra structure is
available\cite{Mat94,Mat95}.  Note also Jackiw's\cite{Jackiw95}
recent demonstration that Dirac quantisation and BRST quantisation
can give {\em different\/} results concerning anomalies: a warning
shot for all those involved in canonical quantum gravity.  At the
very least, the possibility arises that the `true' group in the
quantum theory is some deformation ${\rm Diff}({\cal M})_q$ of the
classical group ${\rm Diff}({\cal M})$.

\subsubsection{B.\ Diffeomorphisms as a partial covariance group}
Another possibility is that the classical group ${\rm Diff}({\cal
M})$ of spacetime diffeomorphisms arises only as a `partial'
covariance group. The two obvious options of this type are:
\begin{itemize}
\item {\em Injective}: ${\rm Diff}({\cal M})$ could appear as a
{\em subgroup\/} of a larger covariance group $G$, as summarised in
the exact sequence
\begin{equation}
	0\rightarrow {\rm Diff}({\cal M})\rightarrow G.
\end{equation}
An example is superstring theory which---at least, in the
perturbative domain---appears to have a much bigger gauge structure
than ${\rm Diff}({\cal M})$ alone.

\item {\em Projective}: ${\rm Diff}({\cal M})$ could be related to a
bigger group $G$ in a projective way: {\em i.e.}, there is some
normal subgroup $K$ of $G$ so that $G/K\simeq \mbox{Diff}({\cal M})$,
as summarised in the exact sequence
\begin{equation}
	0\rightarrow K\rightarrow G\rightarrow \mbox{Diff}({\cal M})
			\rightarrow 0.
\end{equation}
In this case, the relation between the full covariance group $G$ and
the group $\mbox{Diff}({\cal M})$ is analogous to that between $SU2$
and $SO3\simeq SU2/Z_2$, {\em i.e.}, like the relation between
fermions and bosons.
\end{itemize}

\subsection{The Role of Black Holes in Quantum Gravity}
Ever since Hawking's discovery of black hole radiation, a major issue
has been the precise role of black holes in the quantum theory of
gravity. For example, it is tempting to speculate that spacetime at
the Planck scale has a foam-like structure built from `virtual'
black holes. Of course, the use of such language presupposes that
differential geometry is still applicable at this scale, which---as
argued above---is perhaps debatable. However, the idea has many
attractive features and could make sense in a semi-classical
approximation.  Some recent work on the production of virtual
black-hole pairs is reported in Hawking's paper\cite{Haw95} in these
proceedings.

	Another subject of much debate has been the final state of a
collapsing black hole---particularly the fate of the information
that is apparently lost across the event horizon. There are three
different views on this: (i) the information is truly
lost---signalling a fundamental breakdown of conservation of
probability (this is the option that Hawking himself prefers); (ii)
the information is returned in some way in the late stages of the
Hawking radiation; and (iii) the black hole leaves long-lived,
Planck-mass size remnants.

	These ideas are of considerable importance and interest.
However, the feature of black holes on which I wish to focus here is
their possible effect on the nature of quantum physics in a bounded
region. This has been emphasised recently by several people and goes
back to an old remark of Bekenstein\cite{Bek74}: any attempt to
place a quantity of energy $E$ in a spatial region with boundary
area $A$---and such that $E>\sqrt A$---will cause a black hole to
form, and this puts a natural upper bound on the value of the
energy in the region (the argument is summarised nicely in a recent
paper by Smolin\cite{Smo95b}). The implication is that in any theory
of quantum gravity whose semi-classical states contain something
like black-hole backgrounds, the quantum physics of a bounded region
will involve only a {\em finite\/}-dimensional Hilbert space.

	This intriguing possibility is closely related to the so-called
`holographic' hypothesis of t'Hooft\cite{tHoo93} and
Susskind\cite{Sus95} to the effect that physical states in a bounded
region are described by a quantum field theory on the {\em
surface\/} of the region, with a Hilbert space of states that has a
finite dimension---a hypothesis that is itself echoed by recent
ideas in topological quantum field theory, especially the work of
Barrett\cite{Bar95}, Crane\cite{Cra95}, and
Smolin\cite{Smo95a,Smo95b} concerned with the role of topological
quantum field theory in quantum cosmology.

	Ideas of this type could have profound implications for quantum
gravity. In terms of the tower in Figure 1, the implication is that
at one level of phenomenological theory the idea of local spacetime
regions makes sense, and in those regions the quantum theory of
gravity is finite-dimensional. However, in the---presumably
different---tower of phenomenological approximations that includes
weak-field perturbative approaches to quantum gravity, the effective
theory uses an infinite-dimensional Hilbert space to describe the
states of weakly-excited gravitons.

	It is worth remarking that---even in normal quantum theory---an
infinite-dimensional Hilbert space can arise as an approximate
quantisation of a system whose real quantum state space is
finite-dimensional. For example, consider a classical system whose
(compact) phase space is the two-sphere $S^2$. The group $SO3$ acts
as a transitive group of symplectic transformations of this phase
space, and one can argue that quantising the system consists in
finding irreducible representations of this group, all of which---of
course---are finite-dimensional. On the other hand, if one fixes a
point $p\in S^2$ and studies only fluctuations around this point, it
is natural to describe the quantum theory by quantising the system
whose classical state space is the tangent space $\mathR^2$ at $p$.
However, the appropriate group of transformations of $\mathR^2$ is
the familiar Weyl-Heisenberg group of standard wave mechanics, and
the (essentially unique) irreducible representation of this group
has an {\em infinite\/} dimension.

\subsection{Lessons From String Theory}
At this point, it is appropriate to say something about how the
nature of spacetime is seen from the perspective of the various
currently active approaches to quantum gravity.

	The Euclidean programme works mainly with the classical picture
of spacetime as a differentiable manifold. Indeed---in so far as the
formalism does not transcend its own putative semi-classical
limit---the categories used are precisely those of standard general
relativity.

	Similarly, the starting point for canonical quantum gravity is
the structure of a fixed 3-manifold that represents physical space.
Indeed, it is arguable that the use of classical categories is
inevitable in {\em any\/} approach to quantum gravity that is based
on the idea of `quantising' some version of the classical theory of
general relativity. Of course, it is a different question whether or
not the classical picture of space or spacetime is maintained
throughout the development of the theory. For example, the recent
ideas in the Ashtekar programme about the quantisation of area and
volume suggest that the `ultimate' picture of space may have an
essentially discrete aspect.  However, I shall say nothing further
here about such matters, but refer to the comprehensive review by
Ashtekar in the present volume.

	The situation concerning superstring theory is rather different.
True, the perturbative (`$\sigma$-model') approach to string theory
does involve quantising a given classical system, but the system
concerned is not general relativity---and hence the role of space or
spacetime is certainly different from that in the Euclidean or
canonical programmes. However---and perhaps more importantly---there have
been major recent developments in unraveling non-perturbative
aspects of the theory, and these could have dramatic implications
for our understanding of the nature of space and time. We shall
consider perturbative and non-perturbative aspects in turn.

\subsubsection{A.\ Spacetime and perturbative string theory}
Perturbative string theory is based on a `$\sigma$-model' approach
in which the string is viewed as a map $X:{\cal W}\rightarrow {\cal
M}$ from a two-dimensional `world-sheet' $\cal W$ to spacetime $\cal
M$ (the `target' space). The famous Polyakov action for the simplest
theory is
\begin{equation}
S=\int_{{\cal W}} d^2\sigma \sqrt f f^{ab}(\sigma)\partial_aX^\mu(\sigma)
		\partial_bX^\nu(\sigma)\gamma_{\mu\nu}(X(\sigma))
					\label{Polyakov}
\end{equation}
where $f^{ab}$ and $\gamma_{\mu\nu}$ are metric tensors on $\cal W$
and $\cal M$ respectively. This action is conformally invariant, and
to preserve this invariance when $X$ and $f$ are quantised, it
is necessary that (i) the spacetime $\cal M$ has a certain critical
dimension $D$ (the exact value depends on what other fields are
added to the simple bosonic string described by Eq.\
(\ref{Polyakov})); and (ii) the background spacetime metric $\gamma$
satisfies the field equations
\begin{equation}
	R_{\mu\nu}+{\alpha'\over 2}
	R_{\mu\alpha\beta\tau}R_\nu^{\alpha\beta\tau}+O(\alpha'^2)=0
					\label{string-equations}
\end{equation}
where $\alpha'$ is a coupling constant with value around $L_P^2$. It
is by these means that Einstein's equations enter string theory.

	The most relevant observation concerning the nature of spacetime
is that it is still represented by a smooth manifold, although its
dimension may not be four---hence requiring some type of
Kaluza-Klein picture. In addition, it is extremely important to note
that the more realistic superstring theories involve a massless
`dilaton' scalar field $\phi$, and a massless vector particle
described by a three-component field strength $H_{\mu\nu\rho}$,
whose low-energy field equations can be obtained from the effective
spacetime action
\begin{equation}
	S_{\mbox{rm}}=\int d^DX \sqrt\gamma e^{-2\phi}[R-4\gamma^{\mu\nu}
		\partial_\mu\phi\partial_\nu\phi+
			{1\over12}H_{\mu\nu\rho}H^{\mu\nu\rho} 	+O(\alpha')].
\end{equation}
The presence of these extra fundamental fields has a major effect on
the classical solutions of the field equations; in particular, there
have been many studies recently
\footnote{For references, consult the paper by Brandenburger in
these proceedings.} of black-hole and cosmological solutions. The
unavoidable presence of these extra basic fields in superstring
theory---plus the central role of supersymmetry itself---contrasts
sharply with competing approaches in which just the metric
field is quantised.

\subsubsection{B. Duality and the target manifold}
In an action of the type Eq.\ (\ref{Polyakov}), the differential
structure and topology of $\cal M$ are fixed in advance, and there
seems to be no room for any deviation from the classical view of
spacetime. However, there have recently been significant
developments in non-perturbative aspects of string theory, and these
have striking implications for our understanding of the nature of
space and time at the Planck scale.  It is appropriate, therefore,
to say something about this here, even though I can only touch on a
few of the relevant ideas.

	The advances under discussion are based on various types of
`duality' transformations or symmetries: specifically, (i)
$T$-duality; (ii) $S$-duality; and (iii) mirror symmetry. I shall
say a little about each of these in turn.

\medskip
\noindent
{\bf I.\ $T$-duality}:\ The simplest example of target-space duality
(`$T$-duality') arises when the target space is a five-dimensional
manifold of the form ${\cal M}_4\times S^1$. It can be shown that
the physical predictions of the theory are invariant under
replacement of the radius $R$ of the fifth dimension with
$2\alpha'/R$. Thus we cannot differentiate physically between a very
small, and a very large, radius for the additional
dimension---indeed, there is a precise sense in which they are
`gauge' equivalent to each other. This invariance suggests the
existence of a minimum length of $R_{{\rm min}}=\sqrt2\alpha'$, and
can be generalised to more than one extra dimension and with a
topology that is more complex than just a product of circles. This
phenomenon is often cited in support of the claim that strings do
not `see' spacetime in the same way as do point particles: a point
that has been strongly emphasised by Horowitz and collaborators in
their studies of the operational definition of spacetime
singularities in string theory\cite{HS90,HT94}.

\medskip
\noindent
{\bf II.\ $S$-duality}:\ This concept goes back to an old suggestion
by Montonen and Olive\cite{MO77} that the electric-magnetic duality
of source-free electromagnetic theory has an analogue in non-abelian
Yang-Mills theory whereby the physics in the large-coupling limit is
given by the weak-coupling limit of a `dual' theory whose
fundamental entities can be identified with solitonic excitations of
the original theory. The full implementation of this idea requires
the addition of supersymmetry, and it is only relatively recently
that there has been definitive evidence that pairs of such gauge
theories (with $N=2$ supersymmetry) really do arise\cite{SW94}. For
a recent survey see Olive\cite{Oli95}.

	Considerable excitement has been generated recently by the idea
that a similar phenomenon may arise in string theories.  In
particular, there have been well-supported claims\cite{HT95,Wit95a}
that the strong-coupling limit of a type II, ten-dimensional,
superstring theory is a supergravity theory in {\em eleven\/}
dimensions---a duality that is associated naturally with the
introduction of extended objects (`membranes') of dimension greater
than one\cite{DS91}.  Ideas of this type are attractive because (i)
they provide a real possibility of probing the---physically
interesting, but otherwise rather intractable---high-energy limits
of these theories; and (ii) the results suggest that the apparent
plethora of technically viable superstring theories is not as
embarrassingly large as had once been feared: an important step in
securing the overall credibility of the superstring programme.

\medskip
\noindent
{\bf III.\ Mirror symmetry}:\ This is another mechanism whereby
two---apparently very different---string theories can be physically
equivalent. The main developments have been in the context of pairs
of Calabi-Yau manifolds that can be continuously deformed into each
other via an operation involving conifold singularities. From a
physical perspective, these singular points have been identified
with massless black holes\cite{Str95,GMS95}.  The net effect is that
string theories with different compactifications can now be seen as
part of a smoothly connected set, even though the topological
structures of the extra dimensions may be quite different from one
theory to another.
\medskip

	These developments all suggest rather strongly that the
classical ideas of space and time are not applicable at the Planck
length.  Indeed, these new results in string theory are very
compatible with the general idea espoused earlier that spacetime is
not a fundamental category in physics but only something that
applies in a `phenomenological', coarse-grained sense. At a more
technical level, the new ideas suggest that Lagrangian
field-theoretic methods (as represented by the Polyakov action
Eq.\ (\ref{Polyakov})) are reaching the limit of their domain of
applicability and should be replaced by---for example---a more
algebraic approach to theory construction that places less emphasis
on an underlying classical system of fields.

\section{Structural Issues Concerning Quantum Theory}
\subsection{The Key Question}
The question of concern is whether present-day quantum theory can
cope with the demands of quantum gravity. There are several aspects
to this: one is the conceptual problems associated with quantum
cosmology; another concerns the possibility that spacetime itself
should be quantised in some way---an idea that arguably stretches
the current quantum formalism to its limits, both technically and
conceptually.

	It has often been remarked that the instrumentalist tendencies
of the Copenhagen interpretation are inappropriate in quantum
cosmology.  Much to be preferred would be a formalism in which no
fundamental role is ascribed to the idea of `measurement' construed
as an operation external to the domain of the formalism. Of
course---setting aside the needs of quantum cosmology---there has
been extensive debate for many years about finding a more `realist'
interpretation of quantum theory. Two such programmes are the Bohm
approach\cite{Sqi95,CW94,BG95,Val96} to quantum theory, and the
`decoherent histories' approach\cite{Har95}, both of which have been
actively investigated in the last few years for their potential
application to quantum cosmology.

	From a more technical perspective, the main current approaches
to quantum gravity proper---the Euclidean programme, the Ashtekar
scheme, and superstring theory---all use what are, broadly speaking,
standard ideas in quantum theory. In particular---as discussed
above---they work with an essentially classical view of space and
time---something that, arguably, is a {\em prerequisite\/} of the
standard quantum formalism. This raises the important question of
whether quantum theory can be adapted to accommodate the idea that
spacetime itself ({\em i.e.}, not just the metric tensor) is subject
to quantum effects: surely one of the most intriguing challenges to
those working in quantum gravity.

\subsection{Quantising Space-Time}
\label{SSec:QuSpaceTime}
Some of the many issues that arise can be seen by contemplating how
one might try to quantise spacetime `itself' by analogy with what is
done for---say---the simple harmonic oscillator, or the hydrogen
atom. Of course, this may be fundamentally misguided---for example,
the concept of `quantum topology' may be meaningful only in the
coarse-grained sense of belonging to a hierarchy of the type
signified by Figure 1. Nevertheless, it is instructive to think
about the types of problem that occur if one does try to actively
quantise spacetime itself---if nothing else, it reveals the rather
shaky basis of the whole idea of `quantising' a given classical
structure.

	One approach to spacetime quantisation is the `sum over
manifolds' method employed in the Euclidean programme, as in Eq.\
(\ref{Z=sumZ(M)}). Another is to treat spatial topology as some type
of canonical variable\cite{Ish90d}.  And then there are ideas about
using discrete causal sets\cite{Sor91a,Sor91b}, or non-commutative
geometry\cite{Con94} and the like.

	In reflecting on these---and related---schemes for quantising
spacetime, two major issues are seen to arise.  First, a
sophisticated mathematical concept like a differentiable manifold
appears at one end of a hierarchical chain of structure, and it is
necessary to decide at what point in this chain quantum ideas should be
introduced. Second, a given mathematical structure can often be
placed into more than one such chain, and then a decision must be
made about which one to use.

\subsubsection{Two chains leading to $({\cal M},\gamma)$}
For example, if $\gamma$ is a Lorentzian metric on a spacetime
manifold $\cal M$, the pair $({\cal M},\gamma)$ fits naturally into
the chain
\begin{equation}
\mbox{set of spacetime points}\rightarrow\mbox{topology}\rightarrow
\mbox{differential structure}\rightarrow ({\cal M},\gamma)
					\label{chain1}
\end{equation}
where the lowest level ({\em i.e.}, the left hand end) is a set
$\cal M$ of bare spacetime points (with the cardinality of the
continuum), which is then given the structure of a topological space,
which in turn is given the structure of a differentiable manifold
(only possible---of course---for very special topologies) which is
then equipped with a Lorentzian metric to give the final pair
$({\cal M},\gamma)$. Note that a variety of intermediate stages can
be inserted: for example, the link `$\mbox{differential
structure}\rightarrow({\cal M},\gamma)$' could be factored as
\begin{equation}
\mbox{differential structure}\rightarrow
	\mbox{causal structure}\rightarrow({\cal M},\gamma).
\end{equation}

	A quite different scheme arises by exploiting the fact that a
differentiable manifold $\cal M$ is uniquely determined by the
algebraic structure of its commutative ring of differentiable
functions, ${\cal F}({\cal M})$. A ring is a complicated algebraic
structure that can be analysed into a hierarchy of substructures in
several ways. Thus one alternative chain to Eq.\ (\ref{chain1}) is
\begin{equation}
\mbox{set}\rightarrow\mbox{abelian group}\rightarrow
\mbox{vector space}\rightarrow{\cal F}({\cal M})\rightarrow
	({\cal M},\gamma).				\label{chain2}
\end{equation}

\subsection{Three Quantisation Modes}
For any given hierarchical chain that underpins a specific classical
mathematical structure there are at least three different
ways in which quantum ideas might be introduced\cite{Ish94b}.
\begin{enumerate}
	\item {\em Horizontal quantisation}.\ {By this I mean the active
quantisation of one level of the chain whilst keeping fixed all the
structure below. Thus quantum fluctuations occur within a fixed
classical category. For example---in the context of the first chain
above---most approaches to quantum gravity keep fixed the set of
spacetime points, its topology and its differential structure---only
the metric $\gamma$ is quantised (consistent with the fact that, in
{\em classical\/} general relativity, only the metric is a dynamical
variable). One example is the first stage Eq.\ (\ref{Def:Z(M)}) of
the Euclidean programme; another is canonical quantisation of the
3-metric $g_{ab}(x)$ (with space replacing spacetime). More
adventurous would be a scheme in which the set ${\cal M}$ and its
topology are fixed but the differential structure is quantised
\footnote{Percival\cite{Per95} has recently applied the ideas of
`primary state diffusion' to quantise the differential structure of
spacetime.}; or perhaps---as in Eq.\ (\ref{Z=sumZ(M)})---quantum
fluctuations may be restricted to topologies that are compatible
with $\cal M$ being a differentiable manifold.

	Thus, when using the first chain Eq.\ (\ref{chain1}) we are led
naturally to talk of `quantum geometry', `quantum topology', and the
like. However, if applied to the second chain Eq.\ (\ref{chain2}),
quantising within a level leads naturally to considerations of---for
example---the algebraic approach to classical general relativity
pioneered by Geroch\cite{Ger72} (`Einstein algebras') and
non-commutative analogues thereof\cite{PZ95}. Of course, the idea of
a non-commutative version of the ring ${\cal F}({\cal M})$ was one
of the motivating factors behind Connes' seminal ideas on
non-commutative geometry\cite{Con94}.  }

	\item {{\em Trickle-down effects}. This refers to the type of
situation envisaged by Wheeler in his original ideas of quantum
topology in which large quantum fluctuations in a quantised
metric $g_{ab}(x)$ generate changes in the spatial topology.  Thus
active quantisation at one level `trickles down' to produce quantum
effects further down the chain.

	Another example is Penrose's thesis that a projective view of
spacetime structure is more appropriate in quantum gravity, so
that---for example---a spacetime point should be identified with the
collection of all null rays that pass through it. Quantising the
spacetime metric will then induce quantum fluctuations in the null
rays, which will therefore no longer intersect in a single point. In
this way, quantum fluctuations at the top of the first chain
Eq.\ (\ref{chain1}) trickle right down to the bottom of the chain, so
that the very notion of a `spacetime point' acquires quantum
overtones.
	}

	\item {\em Transcendental quantisation}. {From time to time, a
few hardy souls have suggested that a full theory of quantum gravity
requires changing the foundations of mathematics itself. A typical
argument is that standard mathematics is based on set theory, and
certain aspects of the latter (for example, the notion of the
continuum) are grounded ultimately in our spatial perceptions.
However, the latter probe only the world of classical physics, and
hence we feed into the mathematical structures currently used in
{\em all\/} domains of physics, ideas that are essentially classical
in nature. The ensuing category error can be remedied only by
thinking quantum mechanically from the very outset---in other words,
we must look for `quantum analogues' of the categories of standard
mathematics.

	How this might be done
\footnote{Recent examples of this type of
thinking can be found in a book by Finkelstein\cite{Fin95} and a
paper\cite{KF95} of Krause and French.} is by no means obvious. One
approach is to claim that, since classical logic and set theory are
so closely linked (a proposition $P$ determines---and is determined
by---the class of all entities for which $P$ can be rightly
asserted), one should start instead with the formal structure of
{\em quantum\/} logic and try to derive an analogous `non-Boolean
set theory'.  Transcending classical categories in this way is a
fascinating idea, but it is also very iconoclastic
and---career-wise---it is probably unwise to embark on this path before
securing tenure!  }
\end{enumerate}

\subsection{General-Relativity Driven State Reduction}
A very different perspective on the adequacy of standard quantum
theory is given by the idea that the thorny problem of
state-vector reduction itself requires the introduction of general
relativity\cite{KFL86,Pea86,Pen86}.  This position is often
associated with a general view that spacetime is the `ultimate
classical object' and---as such---is not subject to quantum
fluctuations in any way
\footnote{However, Roger Penrose---one of the principal
advocates of this view---has wavered between the idea that (i)
superpositions of spacetimes never occur; and (ii) superpositions
may occur, but they decay in a very short time---a notion that
itself encounters the problem of time (private
communication).}---a position that is diametrically opposite to the
one explored in the previous section.

	This approach to the reduction problem is attractive for several
reasons: (i) gravity is the only {\em universal\/} force we know,
and hence the only force that can be guaranteed to be present in all
physical interactions; and (ii) gravitational effects grow with the
size of the objects concerned---and it is in the context of
macroscopic objects that entangled quantum states are particularly
problematic.

	From a technical perspective, most concrete implementations of
GR-driven state reduction involve variants of the `spontaneous
reduction' theories of the type pioneered by Ghirardi, Rimini and
Weber\cite{GRW86}, and Pearle\cite{Pea89}.  A recent example is the
paper by Pearle and Squires\cite{PS95}, which also contains a good
bibliography.

	As with so many of the other ideas I have discussed in this
article, a key question is whether the notion of gravity-induced
reduction should be built into the theory from the very beginning,
or if it could `emerge' in a phenomenological sense within a tower
of the type in Figure 1.

\section{Where Are We Going?}
I would like to summarise some of the proceeding discussion by
speculating on what a future theory of quantum gravity might look like,
especially in regard to the way it deals with the basic categories
of space and time.

\subsubsection{A.\ Desirable properties of the new theory at a basic level}
At a basic level (or---at least---as high up the tower in Figure 1 as I
am willing to speculate) a future theory of quantum gravity
might have the following features:
\begin{itemize}
\item There will be no fundamental use of the continuum. Applied
in general, this proscribes the use of any set whose cardinality is
greater than countably infinite. Applied in particular, it excludes
a continuum of spacetime points.  Indeed, there should probably be
no fundamental use of the idea of a `spacetime point' at all.

\item At a basic level, the interpretation of the theory must not involve
instrumentalist ideas of the type used in the Copenhagen view; in
particular, there must be no invocation of external
`observers'---conscious or otherwise.

\item {The quantum aspects of the theory will not be grounded in the
use of Hilbert spaces; not even those over a finite field. This
reflects the old idea of Bohr that the wave function of a system
does not refer to the object itself but only to the range of results
that could be obtained by a measurement process for a specific
observable quantity. This view has been resurrected recently by
several authors---in particular, Rovelli\cite{Rov95},
Crane\cite{Cra95}, and Smolin\cite{Smo95a}---in the context of
developing a relational view of quantum theory.  The implication `no
observer, hence no Hilbert space', is not logically inevitable, but
it is one that I find quite attractive.

	As to what should replace Hilbert space, I currently favour the
types of algebraic structure adopted in the past by those working in
quantum logic---especially orthoalgebras and manuals\cite{FGR92}.
This fits in well with an interpretative framework based on the
consistent histories approach to quantum theory\cite{ILS94,Ish95a}.
	}
\end{itemize}

\subsubsection{B.\ Emergent structure in the theory at a
phenomenological level}
In the spirit of Figure 1, let me now list some of the features that
might be expected to emerge from the basic structure in a
phenomenological sense.
\begin{itemize}
\item {\em A continuum spacetime}. At some stage, the familiar
ideas of a continuum spacetime should emerge---perhaps via the
mechanism of an `algebra of local regions' discussed earlier.

\item {\em Standard quantum theory}. {The formalism of standard
quantum theory should also emerge in an appropriate limit. This
would include the usual mathematical framework of Hilbert spaces, but
perhaps augmented with the `holographic hypothesis' that the state
space for physics in a bounded space-time region has a finite
dimension (assuming, of course, that the notions of `space-time
region' and `bounded' make sense at the phenomenological level
concerned).

	I am also very attracted by the idea that state-reduction is
associated with general relativity. However---in the type of theory
being discussed---this would probably be in a `phenomenological'
sense rather than appearing as one of the basic ingredients.  }

\item {\em A theory of quantum gravity}. What we would currently
regard as a `theory of quantum gravity' should also appear at a
phenomenological level once both standard quantum theory and general
relativity have emerged. If our present understanding of quantum
gravity is any guide, this effective quantisation of the
gravitational field will involve a non-local---possibly
string-like---structure.
\end{itemize}

	The last point raises the intriguing question of whether
superstring theory and the loop-variable approach to canonical
quantisation can both be regarded as different modes---or phases---of a
more basic, common structure.  This fascinating possibility is a strong
motivating factor
\footnote{Private communication} behind Smolin's
recent work aimed at relating canonical quantisation and topological
quantum field theory\cite{Smo95a,Smo95b}. A central issue,
presumably, is whether supersymmetry can be assigned some
significant role in the Ashtekar programme.

\subsubsection{C.\ The Key Questions}
It is clear that certain key questions will arise in any attempt to
build a structure of the type envisaged above. Specifically:
\begin{itemize}
\item What is the {\em basic\/} structure (if any) in the theory,
and what emerges as `{\em effective\/}' structure in a more
phenomenological sense?

\item How {\em iconoclastic\/} do we have to be to construct a full theory
of quantum gravity? In particular, is it necessary to go as far as
finding `quantum analogues' of the categories of normal mathematics?

\end{itemize}

	The first question provides a useful way of categorising
potential theoretical frameworks. The second question is the most
basic of all, addressing as it does the challenge of finding the
ingredients for a theory that can head a tower of phenomenological
approximations of the type under discussion. The key problem is to
identify the correct choice of such building blocks among the myriad
of possibilities. This is no easy task, although---as illustrated by
the list of desirable features in a future theory---certain
broad ideas are suggested by the existing research programmes.
Certainly, the momentum behind these approaches---the Ashtekar
programme, superstring theory, and the Euclidean programme---is such
that each is likely to be developed for the foreseeable future,
and---in the process---may yield further ideas for a more radical
approach to the problem of quantum gravity.

	On the other hand, it is possible that none of the current
programmes is on the right track, in which case we need to look
elsewhere for hints on how to proceed.  What is missing, of course,
is any hard {\em empirical\/} data that would enforce a fundamental
shift in approach---which brings us back to a question raised at the
very beginning of the paper: is it possible to find experimental
tests to resolve some of the many obscure issues that cloud the
subject of quantum gravity?

	The obvious problem is the simple dimensional argument
suggesting that effects of quantum gravity will appear at energies
of around $E_P\simeq 10^{-28}\mbox{eV}$, which is well beyond the
range of terrestrial experiments. Of course, there are subtler
possibilities than this. For example:
\begin{itemize}
\item  There may be non-perturbative effects of the type
mentioned earlier in the context of a quantum-gravity induced
ultraviolet cutoff in quantum field theory.

\item Qualitative as well as quantitative
properties of the theory should be considered---some examples of this
type are discussed in a recent paper by Smolin\cite{Smo95c}.

\item There can be unexpected predictions from a theory---for example, the
results reported by Hawking\cite{Haw95} in this volume in his
discussion of virtual black-hole production and its possible
implications for the vanishing $\theta$-angle in QCD.

\item The physics of the immediate post big-bang era may provide an
important testing ground. For example, Hawking has
suggested\cite{Haw93} that the anisotropies in the microwave
background originate in quantum fluctuations around the
Hartle-Hawking ground state\cite{HH85}; while Grischuk\cite{Gri95}
has sought to explain the same phenomena in terms of his work on
squeezed graviton states.
\end{itemize}
So the situation concerning experimental tests is not completely
hopeless. But it is something we must continually strive to improve
if studies in quantum gravity are not to become the 20th century
equivalent of the medieval penchant for computing the cardinality
of angels on pinheads: an ever-present danger at this extreme edge
of modern theoretical physics!

\section*{Acknowledgements}
I am grateful for helpful discussions and correspondence with Jim
Hartle, Gary Horowitz, Karel Kucha{\v r}, Lee Smolin, Kelly Stelle,
Arkady Tseytlin and Steve Weinstein. I am particularly grateful to
Steve Weinstein for his constructive critique of several preliminary
versions of this paper.


\section*{References}
\begin{thebibliography}{10}

\bibitem{Haw75}
S.W. Hawking.
\newblock Particle creation by black holes.
\newblock {\em Comm. Math. Phys.}, 43:199--220, 1975.

\bibitem{Jac95}
E.~Jacobson.
\newblock Thermodynamics of spacetime: {T}he {E}instein equation of state.
\newblock 1995.
\newblock gr-qc/9504004.

\bibitem{BR33}
N.~Bohr and L.~Rosenfeld.
\newblock Zur frage der messbarkeit der elektromagnetischen feldgrossen.
\newblock {\em Kgl.\ Danek Vidensk.\ Selsk.\ Math.-fys.\ Medd.}, 12:8, 1933.

\bibitem{Ros63}
L.~Rosenfeld.
\newblock On quantization of fields.
\newblock {\em Nucl. Phys.}, 40:353--356, 1963.

\bibitem{PG81}
D.N. Page and C.D. Geilker.
\newblock Indirect evidence for quantum gravity.
\newblock {\em Phys. Rev. Lett.}, 47:979--982, 1981.

\bibitem{FH87}
K.~Fredenhagen and R.~Haag.
\newblock Generally covariant quantum field theory and scaling limits.
\newblock {\em Comm. Math. Phys.}, 108:91--115, 1987.

\bibitem{DeW67a}
B.S. DeWitt.
\newblock Quantum theory of gravity. {I}. {T}he canonical theory.
\newblock {\em Phys. Rev.}, 160:1113--1148, 1967.

\bibitem{Kuc92a}
K.~Kucha\v{r}.
\newblock Time and interpretations of quantum gravity.
\newblock In {\em Proceedings of the 4th Canadian Conference on General
  Relativity and Relativistic Astrophysics}, pages 211--314. World Scientific,
  Singapore, 1992.

\bibitem{Ish93}
C.J. Isham.
\newblock Canonical quantum gravity and the problem of time.
\newblock In {\em Integrable Systems, Quantum Groups, and Quantum Field
  Theories}, pages 157--288. Kluwer Academic Publishers, London, 1993.

\bibitem{Dir58b}
P.A.M. Dirac.
\newblock The theory of gravitation in {H}amiltonian form.
\newblock {\em Proc. Royal Soc. of London}, A246:333--343, 1958.

\bibitem{ADM62}
R.~Arnowitt, S.~Deser, and C.W. Misner.
\newblock The dynamics of general relativity.
\newblock In L.~Witten, editor, {\em Gravitation: An Introduction to Current
  Research}, pages 227--265. Wiley, New York, 1962.

\bibitem{Ish92}
C.J. Isham.
\newblock Conceptual and geometrical problems in quantum gravity.
\newblock In H.~Mitter and H.~Gausterer, editors, {\em Recent Aspects of
  Quantum Fields}, pages 123--230. Springer-Verlag, Berlin, 1992.

\bibitem{Kuc93}
K.~Kucha\v{r}.
\newblock Canonical quantum gravity.
\newblock In R.J. Gleiser, C.N. Kozameh, and O.M. Moreschi, editors, {\em
  General Relativity and Gravitation, 1992}, pages 119--150. IOP Publishing,
  Bristol, 1993.

\bibitem{Whe64}
J.A. Wheeler.
\newblock Geometrodynamics and the issue of the final state.
\newblock In C.~DeWitt and B.S. DeWitt, editors, {\em Relativity, Groups and
  Topology}, pages 316--520. Gordon and Breach, New York and London, 1964.

\bibitem{Whe68}
J.A. Wheeler.
\newblock Superspace and the nature of quantum geometrodynamics.
\newblock In C.~DeWitt and J.W. Wheeler, editors, {\em Batelle Rencontres: 1967
  Lectures in Mathematics and Physics}, pages 242--307. Benjamin, New York,
  1968.

\bibitem{Mis69a}
C.W. Misner.
\newblock Quantum cosmology {I}.
\newblock {\em Phys. Rev.}, 186:1319--1327, 1969.

\bibitem{Eve57}
H.~Everett.
\newblock Relative state formulation of quantum mechanics.
\newblock {\em Rev. Mod. Phys.}, 29:141--149, 1957.

\bibitem{Whe57}
J.A. Wheeler.
\newblock Assessment of {E}verett's ``relative state'' formulation of quantum
  theory.
\newblock {\em Rev. Mod. Phys.}, 29:463--465, 1957.

\bibitem{Pen67}
R.~Penrose.
\newblock Twistor theory.
\newblock {\em J. Math. Phys.}, 8:345--366, 1967.

\bibitem{Fey63}
R.~Feynman.
\newblock {\em Acta Physical Polonica}, XXIV:697, 1963.

\bibitem{DeW65}
B.S. DeWitt.
\newblock {\em Dynamical Theory of Groups and Fields}.
\newblock Wiley, New York, 1965.

\bibitem{DeW67b}
B.S. DeWitt.
\newblock Quantum theory of gravity. {II}. {T}he manifestly covariant theory.
\newblock {\em Phys. Rev.}, 160:1195--1238, 1967.

\bibitem{Man68b}
S.~Mandlestan.
\newblock Feynman rules for the gravitational field from the
  coordinate-independent field theoretic formalism.
\newblock {\em Phys. Rev.}, 175:1604--1623, 1968.

\bibitem{FP67}
L.~Fadeev and V.~Popov.
\newblock Feynman rules for {Yang-Mills} theory.
\newblock {\em Phys. Lett.}, 25B:27--30, 1967.

\bibitem{SW64}
R.F. Streater and A.S. Wightman.
\newblock {\em PCT, Spin and Statistics, and All That}.
\newblock Benjamin, New York, 1964.

\bibitem{IPS81}
C.J. Isham, R.~Penrose, and D.W. Sciama.
\newblock {\em Quantum Gravity: {A} Second {O}xford Symposium}.
\newblock Clarendon Press, 1981.

\bibitem{GP78}
G.W. Gibbons and M.J. Perry.
\newblock Black holes and thermal {G}reen functions.
\newblock {\em Proc. Royal Soc. of London}, A358:467--494, 1987.

\bibitem{GH93}
M.~Gell-{M}ann and J.~Hartle.
\newblock Classical equations for quantum systems.
\newblock {\em Phys. Rev.}, D47:3345, 1993.

\bibitem{HH83}
J.B. Hartle and S.W. Hawking.
\newblock Wave function of the universe.
\newblock {\em Phys. Rev.}, D28:2960--2975, 1983.

\bibitem{IPS75}
C.J. Isham, R.~Penrose, and D.W. Sciama.
\newblock {\em Quantum Gravity: {A}n {O}xford Symposium}.
\newblock Clarendon Press, 1975.

\bibitem{GS86}
M.H. Goroff and A.~Sagnotti.
\newblock The ultraviolet behavour of {E}instein gravity.
\newblock {\em Nucl. Phys.}, B266:709--736, 1986.

\bibitem{Con94}
A.~Connes.
\newblock {\em Non Commutative Geometry}.
\newblock Academic Press, New York, 1994.

\bibitem{Dje95}
A.E.F. Djemai.
\newblock Introduction to {Dubois-Violette}'s noncommutative differential
  geometry.
\newblock {\em Int. J. Theor. Phys.}, 34:801--887, 1995.

\bibitem{Gib95}
P.~Gibbs.
\newblock The small-scale structure of space-time: {A} bibliographical review.
\newblock 1995.
\newblock hep-th/9506171.

\bibitem{vN81}
P.~{van Nieuwenhuizen}.
\newblock Supergravity.
\newblock {\em Physics Reports}, 68:189--398, 1981.

\bibitem{Pru95}
E.~Prugovecki.
\newblock {\em Principles of Quantum General Relativity}.
\newblock World Scientific, Singapore, 1995.

\bibitem{Gar95}
L.J. Garay.
\newblock Quantum gravity and minimum length.
\newblock {\em Int. Jour. Mod. Phys. A}, 10:145--165, 1995.

\bibitem{Ash86}
A.~Ashtekar.
\newblock New variables for classical and quantum gravity.
\newblock {\em Phys. Rev. Lett.}, 57:2244--2247, 1986.

\bibitem{RS90}
C.~Rovelli and L.~Smolin.
\newblock Loop space representation of quantum general relativity.
\newblock {\em Nucl. Phys.}, B331:80--152, 1990.

\bibitem{Car95a}
S.~Carlip.
\newblock Lectures on {$(2+1)$}-dimensional gravity.
\newblock 1995.
\newblock Lectures given at the First Seoul Workshop on Gravity and Cosmology,
  February 1995; gr-qc/9503024.

\bibitem{AS91}
A.~Ashtekar and J.~Stachel.
\newblock {\em Conceptual Problems of Quantum Gravity}.
\newblock Birkh{\"a}user, Boston, 1991.

\bibitem{Bra95}
R.H. Brandenberger.
\newblock Nonsingular cosmology and {P}lanck scale physics.
\newblock In {\em Proceedings of the International Workshop on Planck Scale
  Physics, India 1994}. World Scientific, Singapore, 1995.

\bibitem{FT81}
T.S. Fradkin and A.A. Tseytlin.
\newblock Renormalizable asymptotically free quantum theory of gravity.
\newblock {\em Phys. Lett.}, 104B:377--381, 1981.

\bibitem{Tse95}
A.A. Tseytlin.
\newblock Black holes and exact solutions in string theory.
\newblock In {\em Proceedings of the International School of Astrophysics;
  Erice, September 1994}. World Scientific, Singapore, 1995.
\newblock hep-th/9410008.

\bibitem{Sor83}
R.D. Sorkin.
\newblock Posets as lattice topologies.
\newblock In B.~Bertotti, F.~de~Felice, and A.~Pascolini, editors, {\em General
  Relativity and Gravitation: Proceedings of the {GR10} Conference. {V}olume
  {I}}, pages 635--637. Consiglio Nazionale Delle Ricerche, Rome, 1983.

\bibitem{Sor91a}
R.D. Sorkin.
\newblock Finitary substitute for continuous topology.
\newblock {\em Int. J. Theor. Phys.}, 30:923--947, 1991.

\bibitem{Ish90d}
C.J. Isham.
\newblock An introduction to general topology and quantum topology.
\newblock In H.C. Lee, editor, {\em Physics, Geometry and Topology}, pages
  129--190. Plenum Press, New York, 1990.

\bibitem{Mat94}
{H-J}. Matschull.
\newblock On loop states in quantum gravity and supergravity.
\newblock {\em Class. Quan. Grav.}, 11:2395--2410, 1994.

\bibitem{Mat95}
{H-J}. Matschull.
\newblock New representation and a vacuum state for canonical quantum gravity.
\newblock {\em Class. Quan. Grav.}, 12:651--676, 1995.

\bibitem{Jackiw95}
R.~Jackiw.
\newblock Quantum modifications to the {W}heeler-{D}e{W}itt equation.
\newblock 1995.
\newblock gr-qc/9506037.

\bibitem{Haw95}
S.W. Hawking.
\newblock Virtual black holes.
\newblock 1995.
\newblock hep-th/9510029.

\bibitem{Bek74}
J.D. Bekenstein.
\newblock The quantum mass spectrum of a {K}err black hole.
\newblock {\em Lett.\ Nuov.\ Cim.}, 11:467--470, 1974.

\bibitem{Smo95b}
L.~Smolin.
\newblock The {B}ekenstein bound, topological quantum field theory and
  pluralistic quantum cosmology.
\newblock 1995.
\newblock gr-qc/9508064.

\bibitem{tHoo93}
G.~t'Hooft.
\newblock Dimensional reduction in quantum gravity.
\newblock 1993.
\newblock gr-qc/9310006.

\bibitem{Sus95}
L~Susskind.
\newblock The world as a hologram.
\newblock {\em J. Math. Phys.}, 1995.
\newblock hep-th/9409089.

\bibitem{Bar95}
J.W. Barrett.
\newblock Quantum gravity as topological quantum field theory.
\newblock {\em J. Math. Phys.}, 1995.
\newblock to appear; gr-qc/9506070.

\bibitem{Cra95}
L.~Crane.
\newblock Clocks and categories: {I}s quantum gravity algebraic?
\newblock {\em J. Math. Phys.}, 1995.
\newblock to appear; gr-qc/9504038.

\bibitem{Smo95a}
L.~Smolin.
\newblock Linking topological quantum field theory and nonperturbative quantum
  gravity.
\newblock {\em J. Math. Phys.}, 1995.
\newblock gr-qc/9505028.

\bibitem{HS90}
G.~Horowitz and A.~Steif.
\newblock Spacetime singularities in string theory.
\newblock {\em Phys. Rev. Lett.}, 64:260--263, 1990.

\bibitem{HT94}
G.~Horowitz and A.A. Tseytlin.
\newblock Exact solutions and singularities in string theory.
\newblock {\em Phys. Rev.}, D30:5204--5224, 1994.

\bibitem{MO77}
C.~Montonen and D.I. Olive.
\newblock Magnetic monopoles as gauge particles?
\newblock {\em Phys. Lett.}, B72:117--120, 1977.

\bibitem{SW94}
N.~Seiberg and E.~Witten.
\newblock Electric-magnetic duality, monopole condensation, and confinement in
  {$N=2$} supersymmetric {Y}ang-{M}ills theory.
\newblock {\em Nucl. Phys.}, B426:19--52, 1994.

\bibitem{Oli95}
D.I. Olive.
\newblock Exact electromagnetic duality.
\newblock In {\em Proceedings of the Trieste Conference on Recent Developments
  in Statistical Mechanics and Quantum Field Theory, April 1995}, 1995.
\newblock hep-th/9508089.

\bibitem{HT95}
C.M. Hull and P.K. Townsend.
\newblock Unity of superstring dualities.
\newblock {\em Nucl. Phys.}, B438:109--137, 1995.

\bibitem{Wit95a}
E.~Witten.
\newblock String theory dynamics in various dimensions.
\newblock {\em Nucl. Phys.}, B443:85--126, 1995.

\bibitem{DS91}
M.J. Duff and K.S. Stelle.
\newblock Multimembrane solutions of {$D=11$} supergravity.
\newblock {\em Phys. Lett.}, B253:113--118, 1991.

\bibitem{Str95}
A.~Strominger.
\newblock Massless black holes and conifolds in string theory.
\newblock {\em Nucl. Phys.}, B451:96--108, 1995.

\bibitem{GMS95}
B.R. Greene, D.R. Morrison, and A.~Strominger.
\newblock Black hole condensation and the unification of string vacua.
\newblock {\em Nucl. Phys.}, B451:109--120, 1995.

\bibitem{Sqi95}
E.J. Squires.
\newblock Quantum theory, relativity, and the {B}ohm model.
\newblock {\em Annals of New York Academy of Sciences}, 755:451--463, 1995.

\bibitem{CW94}
C.~Callender and R.~Weingard.
\newblock The {B}ohmian model of quantum cosmology.
\newblock In D.~Hull, M.~Forbes, and R.M. Burian, editors, {\em PSA 1994:
  Proceedings of the 1994 Biennial Meeting of the Philosophy of Science
  Association; Volume One}. Philosophy of Science Association, East Lansing,
  Michigan, 1994.

\bibitem{BG95}
A.~Blaut and J.K. Glikman.
\newblock Quantum potential approach to quantum cosmology.
\newblock 1995.
\newblock gr-qc/9509040; to appear in {{\em Class. Qu. Grav}}.

\bibitem{Val96}
A.~Valentini.
\newblock {\em On the Pilot-Wave Theory of Classical, Quantum and Subquantum
  Physics}.
\newblock Springer-Verlag, Berlin, 1996.

\bibitem{Har95}
J.~Hartle.
\newblock Spacetime quantum mechanics and the quantum mechanics of spacetime.
\newblock In B.~Julia and J.~Zinn-Justin, editors, {\em Proceedings on the 1992
  Les Houches School, Gravitation and Quantisation}, pages 285--480. Elsevier
  Science, 1995.

\bibitem{Sor91b}
R.D. Sorkin.
\newblock Spacetime and causal sets.
\newblock In J.C. D'Olivo, E.~Nahmad-Achar, M.~Rosenbaum, M.P. Ryan, L.F.
  Urrutia, and F.~Zertuche, editors, {\em Relativity and Gravitation: Classical
  and Quantum}, pages 150--173. World Scientific, Singapore, 1991.

\bibitem{Ish94b}
C.J. Isham.
\newblock Prima facie questions in quantum gravity.
\newblock In J.~Ehlers and H.~Friedrich, editors, {\em Canonical Relativity:
  Classical and Quantum}, pages 1--21. Springer-Verlag, Berlin, 1994.

\bibitem{Per95}
I.C. Percival.
\newblock Quantum space-time fluctuations and primary state diffusion.
\newblock {\em Proc. Royal Soc. of London}, 1995.
\newblock to appear; quant-ph/9508021.

\bibitem{Ger72}
R.~Geroch.
\newblock Einstein algebras.
\newblock {\em Comm. Math. Phys.}, 25:271--275, 1972.

\bibitem{PZ95}
G.N. Parfionov and R.R. Zapatrin.
\newblock Pointless spaces in general relativity.
\newblock 1995.
\newblock gr-qc/9503048.

\bibitem{Fin95}
D.R. Finkelstein.
\newblock {\em Quantum Relativity}.
\newblock Springer-Verlag, Berlin, 1995.

\bibitem{KF95}
D.~Krause and S.~French.
\newblock A formal framework for quantum individuality.
\newblock {\em Synthese}, 102:195--214, 1995.

\bibitem{KFL86}
F.~K\'arolyh\'azy, A.~Frenkel, and B.~Luk\'acs.
\newblock On the possible role of gravity in the reduction of the state vector.
\newblock In R.~Penrose and C.J. Isham, editors, {\em Quantum Concepts in Space
  and Time}, pages 109--128. Clarendon Press, Oxford, 1986.

\bibitem{Pea86}
P.~Pearle.
\newblock Models for reduction.
\newblock In R.~Penrose and C.J. Isham, editors, {\em Quantum Concepts in Space
  and Time}, pages 84--108, Oxford, 1986. Clarendon Press.

\bibitem{Pen86}
R.~Penrose.
\newblock Gravity and state vector reduction.
\newblock In R.~Penrose and C.J. Isham, editors, {\em Quantum Concepts in Space
  and Time}, pages 129--146. Clarendon Press, Oxford, 1986.

\bibitem{GRW86}
G.C. Ghirardi, A.~Rimini, and T.~Weber.
\newblock Unified dynamics for microscopic and macroscopic systems.
\newblock {\em Phys. Rev.}, D34:470--491, 1986.

\bibitem{Pea89}
P.~Pearle.
\newblock Combining stochastical dynamical state-vector reduction with
  spontaneous localization.
\newblock {\em Phys. Rev.}, A39:2277--2289, 1989.

\bibitem{PS95}
P.~Pearle and E.~Squires.
\newblock Gravity, energy conservation and parameter values in collapse models.
\newblock 1995.
\newblock quant-ph/9503019.

\bibitem{Rov95}
C.~Rovelli.
\newblock Relative quantum theory.
\newblock 1995.
\newblock hep-th/9403015.

\bibitem{FGR92}
D.J. Foulis, R.J. Greechie, and G.T. {R\"uttimann}.
\newblock Filters and supports in orthoalgebras.
\newblock {\em Int. J. Theor. Phys.}, 31:789--807, 1992.

\bibitem{ILS94}
C.J. Isham, N.~Linden, and S.~Schreckenberg.
\newblock The classification of decoherence functionals: an analogue of
  {G}leason's theorem.
\newblock {\em J. Math. Phys.}, 35:6360--6370, 1994.

\bibitem{Ish95a}
C.J. Isham.
\newblock Quantum logic and decohering histories.
\newblock In T.~Tchrakian, editor, {\em Theories of Fundamental Interactions}.
  World Scientific Press, Singapore, 1995.

\bibitem{Smo95c}
L.~Smolin.
\newblock Experimental signatures of quantum gravity.
\newblock 1995.
\newblock gr-qc/9503027.

\bibitem{Haw93}
S.W. Hawking.
\newblock {The Nature of Space and Time}.
\newblock 1993.
\newblock Lectures given at the Newton Institute, Cambridge.

\bibitem{HH85}
J.J Halliwell and S.W. Hawking.
\newblock Origin of structure in the universe.
\newblock {\em Phys. Rev.}, D31:1777--1791, 1985.

\bibitem{Gri95}
L.P. Grischuk.
\newblock Statistics of the microwave background anisotropies caused by the
  squeezed cosmological perturbations.
\newblock 1995.
\newblock gr-qc/9504045.

\end{thebibliography}
\end{document}